\documentclass[12pt]{article}

\usepackage[makeroom]{cancel}
\usepackage[normalem]{ulem}
\usepackage{amsmath,amssymb}
\usepackage{slashed}
\usepackage{xcolor}
\usepackage{setspace}
\usepackage{graphicx}
\usepackage{cite}
\usepackage{pdflscape}
\usepackage{multirow}
\usepackage[thinlines]{easytable}
\usepackage{url}
\usepackage{braket}
\usepackage{subfigure}
\usepackage{longtable}
\usepackage{booktabs}
\usepackage[colorinlistoftodos]{todonotes}

\newcommand{\fref}[1]{Fig.~\ref{fig:#1}} 
\newcommand{\eref}[1]{Eq.~\eqref{eq:#1}}

\newcommand{\aref}[1]{Appendix~\ref{app:#1}}
\newcommand{\sref}[1]{Section~\ref{sec:#1}}
\newcommand{\cref}[1]{Chapter~\ref{ch:#1}}

\newcommand{\nnl}{\nonumber \\}

\newcommand{\beq}{\begin{equation}} 
\newcommand{\eeq}{\end{equation}} 
\newcommand{\ba}{\begin{array}}  
\newcommand{\ea}{\end{array}} 
\newcommand{\bea}{\begin{eqnarray}}  
\newcommand{\eea}{\end{eqnarray} }  
\newcommand{\be}{\begin{eqnarray}}  
\newcommand{\ee}{\end{eqnarray} }  
\newcommand{\bal}{\begin{align}}
\newcommand{\eal}{\end{align}}   
\newcommand{\bi}{\begin{itemize}}  
\newcommand{\ei}{\end{itemize}}  
\newcommand{\ben}{\begin{enumerate}}  
\newcommand{\een}{\end{enumerate}}  
\newcommand{\bc}{\begin{center}}
\newcommand{\ec}{\end{center}} 
\newcommand{\bt}{\begin{table}}
\newcommand{\et}{\end{table}}  
\newcommand{\btb}{\begin{tabular}}
\newcommand{\etb}{\end{tabular}}

\newcommand{\cO}{{\mathcal O}}

\newcommand{\cM}{{\mathcal M}}

\newcommand{\re}{{\mathrm{Re}} \,}
\newcommand{\im}{{\mathrm{Im}} \,}

\def\hc{{\rm h.c.}} 

\newcommand{\eps}{\epsilon}

\begin{document}

\begin{titlepage}

\begin{flushright}
CERN-TH-2019-002\\
LPT Orsay 19-02 \\
\end{flushright}

\begin{center}
\begin{spacing}{1.5}
{\LARGE \bf
Reactor neutrino oscillations \\
 as constraints on Effective Field Theory
}\end{spacing} %
\vspace{1.4cm}

\renewcommand{\thefootnote}{\fnsymbol{footnote}}
{Adam~Falkowski$^a$, Mart\'{i}n~Gonz\'{a}lez-Alonso$^{b}$, and Zahra~Tabrizi$^{c}$}
\renewcommand{\thefootnote}{\arabic{footnote}}
\setcounter{footnote}{0}

\vspace*{.8cm}
\centerline{${}^a$\it Laboratoire de Physique Th\'{e}orique (UMR8627), CNRS, Univ. Paris-Sud,}
\centerline{\it Universit\'{e} Paris-Saclay, 91405 Orsay, France}\vspace{1.3mm}
\centerline{${}^b$\it Theoretical Physics Department, CERN, 1211 Geneva 23, Switzerland}\vspace{1.3mm}
\centerline{${}^c$\it Departamento de F\'isica Matem\'atica, Instituto de F\'isica, }
\centerline{\it Universidade de S\~ao Paulo, C. P. 66.318, 05315-970 S\~ao Paulo, Brazil}\vspace{1.3mm}
\vspace{.2cm}
\it{E-mail:~} adam.falkowski@th.u-psud.fr,
martin.gonzalez.alonso@cern.ch,
ztabrizi@if.usp.br

\vspace*{.2cm}

\end{center}

\vspace*{10mm}
\begin{abstract}\noindent\normalsize

We study constraints on the Standard Model Effective Field Theory (SMEFT) from neutrino oscillations in short-baseline reactor experiments. 
We calculate the survival probability of reactor antineutrinos at the leading order in the SMEFT expansion, that is including linear effects of dimension-6 operators.  
It is shown that, at this order, reactor experiments alone cannot probe  charged-current contact interactions between leptons and quarks that are of the (pseudo)vector (V$\pm$A) or pseudo-scalar type.
We also note that flavor-diagonal (pseudo)vector coefficients do not have observable effects in oscillation experiments. 
In this we reach novel or different conclusions than prior analyses of non-standard neutrino interactions. 
On the other hand, reactor experiments offer a unique opportunity to probe tensor and scalar SMEFT operators that are off-diagonal in the lepton-flavor space.
We derive constraints on the corresponding SMEFT parameters using the most recent data from the Daya Bay and RENO experiments.

\end{abstract}

\end{titlepage}
\newpage 

\renewcommand{\theequation}{\arabic{section}.\arabic{equation}} 

\tableofcontents

\section{Introduction}
\label{sec:intro}

Since the discovery of neutrino oscillations, the field of precision neutrino physics has experienced a formidable rate of progress. 
Assuming the standard 3-flavor picture, the mass squared differences between the neutrino eigenstates and all three angles in the mixing matrix have been determined with a good precision, see Ref.~\cite{Esteban:2018azc} for a recent update.  
The standard parameters are now overconstrained by multiple independent  measurements, with overall a good consistency. 
In a way, the situation is similar to that in electroweak precision physics in the 1990s when, given the wealth of precise and theoretically clean information from LEP-1, the initial focus on measuring the parameters of the Standard Model (SM)
could be extended to constraining hypothetical phenomena (technicolor, supersymmetry, etc.).
By the same token, neutrino experiments now have a potential to systematically explore new physics beyond the neutrino masses and mixings. 

One such area of exploration are the so-called non-standard interactions (NSI). 
Oscillation experiments are sensitive not only to neutrino masses and mixing, but also to how neutrinos interact with matter. 
The SM makes precise predictions about these interactions, which however can be perturbed by physics beyond the SM (BSM). 
In particular, new effective 4-fermion interactions between leptons and quarks may give observable effects in neutrino production, propagation, and detection, and thus they can be constrained by experiment.  
These studies have a long history in the literature, see e.g.~\cite{Antusch:2006vwa,Kopp:2007ne,Bolanos:2008km,Ohlsson:2008gx,Delepine:2009am,Biggio:2009nt,Leitner:2011aa,Ohlsson:2012kf,Esmaili:2013fva,Agarwalla:2014bsa,Ludl:2016ane,Choudhury:2018xsm,Heeck:2018nzc,Altmannshofer:2018xyo,AristizabalSierra:2018eqm,Coloma:2017ncl,Esteban:2018ppq,Bergmann:1999rz} and \cite{Farzan:2017xzy} for a recent review. 

Previous works were contended with an ad-hoc parametrization of NSI, using the effective couplings $\epsilon^{s,d,m}$ to describe the non-standard effects in production, detection, and propagation of neutrino states (see \aref{NSInu} for more details).
There has been little emphasis on connecting these couplings to parameters of concrete BSM models, or to Wilson coefficients of a well-defined and systematic framework of effective field theory (EFT). 
Consequently, full attention has not been paid to  such issues as  power counting of NSI effects, extraction of the mixing angles in the presence of general new physics, or comparison between the sensitivity of oscillation and other experiments. 
We argue here that there are distinct advantages in embedding NSI in a solid EFT.  
First, consistent EFTs come with an expansion parameter, and the Lagrangian, amplitudes, and observables can be systematically constructed order by order in that expansion. 
This allows one to compare different NSI effects in neutrino oscillations, and unambiguously identify the leading order contributions.    
Moreover, EFTs may predict correlations between the magnitude of effects in neutrino oscillation and in other precision experiments, such as nuclear beta transitions, meson decays, Drell-Yan production at the LHC, etc.  
In this picture, oscillation experiments become an ingredient in the broad program of precision measurements. 
Moreover, sensitivities of the different precision probes can be meaningfully compared. 

In this paper we propose a systematic EFT approach to neutrino oscillations. We focus on short-baseline reactor neutrino experiments, however the formalism can be readily applied to experiments with longer baselines and for different neutrino production and detection processes. 
Our point of departure is the so-called SMEFT, where higher-dimensional  interactions invariant under the local $SU(3)_C \times SU(2)_L \times U(1)_Y$ symmetry are added to the SM. 
They are organized in an expansion in $1/\Lambda$, where $\Lambda$ can be interpreted as the BSM scale suppressing the higher-dimensional operators. 
Given the SMEFT Lagrangian, we derive the effective charged-current interactions between neutrinos, charged leptons, and nucleons in the low-energy EFT relevant for reactor experiments. 
This allows us to calculate the survival probability of an electron antineutrino in short-baseline experiments in the presence of general dimension-6 (order $\Lambda^{-2}$) SMEFT interactions.   
We point out that these experiments offer a unique opportunity to probe, at the linear level, certain SMEFT operators that are off-diagonal in the lepton-flavor space.
We identify the linear combinations of Wilson coefficients that are constrained by reactor experiments at the leading order in the SMEFT expansion. 
We then proceed to obtain numerical constraints on these combinations using the most recent  data from the Daya Bay~\cite{Adey:2018zwh} and RENO~\cite{Bak:2018ydk} experiments. 

Our systematic approach puts into perspective some of the conclusions reached in the prior NSI literature.
We will argue that, at the leading order in the SMEFT expansion, NSI interactions {\em diagonal} in the lepton-flavor space are currently {\em not} probed by oscillation experiments.  
More precisely, any modifications of diagonal V$\pm$A interactions are fully absorbed in the phenomenological extraction of SM parameters (the CKM element $V_{ud}$ and the neutron axial charge $g_A$), while for the scalar and tensor ones stringent model-independent constraints from nuclear beta decays exclude observable signals in reactor experiments, given the current precision of the latter.
As for NSI {\em off-diagonal} in the lepton-flavor indices, those involving  only left-handed leptons and quarks (V-A type) cannot be constrained by the reactor experiments alone, as they merely renormalize the a-priori unknown mixing angle $\theta_{13}$. 
Off-diagonal NSI with right-handed quarks (V+A) are in principle observable in the reactor oscillation experiments, however they do not arise at $\cO(\Lambda^{-2})$ from  dimension-6 SMEFT operators. 
On the other hand, reactor experiments show an interesting sensitivity to off-diagonal tensor and scalar NSI, 
which were actually neglected in most prior studies.

This paper has the following structure. 
In \sref{eft} we review the formalism of the SMEFT and the resulting EFT below the weak scale. 
In \sref{oscillations} we derive the dependence on the SMEFT parameters  of the anti-neutrino survival probability in reactor experiments. 
The constraints on these parameters from the Daya Bay and RENO observations are presented in \sref{fit}, and compared in \sref{non-osc} to the constraints from other precision experiments.
We summarize our findings in \sref{conclusion} and comment on the significance of our results on the NSI program in oscillation experiments.

\section{EFT ladder}
\setcounter{equation}{0}
\label{sec:eft}

As mentioned above, the oscillation pattern of neutrinos depends not only on their mass differences, but also on their interactions with other particles.
We are interested in the situation where these interactions deviate from the SM predictions. 
Our goal is to derive new constraints on fundamental theories with  heavy BSM particles, without referring to a specific model. 
For that reason we will use the language of EFT.  
In this section we review the crucial elements of EFTs relevant for our analysis. 

\subsection{SMEFT}

If new particles beyond the SM are much heavier than the $Z$ boson and the electroweak symmetry breaking is linearly realized, then the relevant effective theory above the weak scale is the so-called SMEFT~\cite{Buchmuller:1985jz,Grzadkowski:2010es}. 
It has the same local symmetry and particle content as the SM, which in particular entails the absence of right-handed neutrinos. But the SMEFT differs from the SM by the presence of higher-dimensional (non-renormalizable) interactions in the Lagrangian, which provide an effective description of physical effects of heavy BSM particles. 
They are organized in a systematic expansion in operator dimensions, with each consecutive terms suppressed by a higher power of the new physics scale $\Lambda$. 
Dimension-5 operators are essential as they give rise to Majorana masses of the SM neutrinos. 
The formulas presented in this work assume the normal ordering of neutrino masses, but the changes in the case of inverse ordering are trivial, as we will explicitly discuss. 
Due to the smallness  of the neutrino masses, dimension-5 operators have negligible effects on production and detection amplitudes of relativistic neutrinos.      
On the other hand, these can be significantly affected by dimension-6 operators suppressed by $\Lambda^{-2}$. 
In particular, some dimension-6 operators lead to deviations of the couplings of the SM quarks and leptons to the $W$ boson from the SM prediction; others introduce new contact interactions between quarks and leptons. 
In our study  we will ignore the effects of operators with dimensions higher than six, which are suppressed by more than two powers of $\Lambda$. Consequently, we will only trace new physics corrections linear (order $\Lambda^{-2}$) in Wilson coefficients of dimension-6 operators, and ignore the quadratic effects that are $\cO(\Lambda^{-4})$.

\subsection{WEFT}

The SMEFT is a convenient tool when it comes to studying high-energy physics above the weak scale.  
However, neutrino oscillation experiments are performed at energies well below the weak scale. 
At the scale $\mu \lesssim m_W$, the $W$ and $Z$ bosons, as well as the Higgs boson and the top quark, can  be integrated out from the SMEFT, leading to another effective theory that we refer to as the weak EFT (WEFT).\footnote{In this work we consider the WEFT as the low-energy theory of the SMEFT. However, the WEFT is a consistent EFT in its own right, which can be valid even if it is not completed by the SMEFT at higher energies. See \aref{without} for the discussion of such a set-up.}
It has a smaller particle content and different interactions than the SMEFT. 
Below we focus on the charged-current 4-fermion interactions between the up and down quarks and the 3 generations of charged leptons and neutrinos. 
At the leading order in the WEFT we can parametrize them as
\bea 
\label{eq:EFT_lweft}
\cal L_{\rm WEFT} 
& \supset &
- \,\frac{2 V_{ud}}{v^2} \big \{   
\left [ {\bf 1} +    \epsilon_L \right]_{\alpha \beta}  (\bar u  \gamma^\mu P_L d)  (\bar \ell_\alpha  \gamma_\mu P_L \nu_\beta) 
 +  [\epsilon_R]_{\alpha \beta}  (\bar u  \gamma^\mu P_R d)   (\bar \ell_\alpha  \gamma_\mu P_L \nu_\beta)
\nnl 
&&
~~~~~~~~~~+\, {1 \over 2 } [\epsilon_S]_{\alpha \beta} (\bar u d) (\bar \ell_\alpha P_L  \nu_\beta)  
- {1 \over 2} [\epsilon_P]_{\alpha \beta} (\bar u \gamma_5 d) (\bar \ell_\alpha P_L  \nu_\beta)   
\nnl 
&&
~~~~~~~~~~+\,  {1 \over 4} [\hat \epsilon_T]_{\alpha \beta} (\bar u  \sigma^{\mu \nu} P_L d)   (\bar \ell_\alpha  \sigma_{\mu \nu} P_L \nu_\beta) 
+ \hc  \big \}~,
\eea
where $v$ is the VEV of the Higgs doublet, $V_{ud}$ is a CKM matrix element, $\ell_\alpha = e,\mu,\tau$ is a charged lepton field, $\sigma^{\mu\nu}=i[\gamma^\mu,\gamma^\nu]/2$, and $P_{L,R}$ are the usual chirality projectors $(1\mp\gamma_5)/2$.\footnote{
The hat over $\hat \epsilon_T$ indicates that the normalization differs by a factor of 4 from $\epsilon_T$ used e.g. in~\cite{Falkowski:2017pss}. 
The present normalization is more natural in the sense that typical new physics models generating tensor interactions will give comparable contribution to $\hat \epsilon_T$ and $\epsilon_{S,P}$, see e.g. \cite{deBlas:2017xtg}. 
}
Above, the fields $u$, $d$, $\ell_\alpha$ are written in the basis where their mass terms are diagonal. 
The  flavor neutrino states $\nu_\alpha$ are connected to the mass eigenstates by $\nu_{\alpha} = U_{\alpha J} \nu_J$, where $\alpha = e,\mu,\tau$, $J=1,2,3$, and $U$ is the unitary PMNS matrix parametrized by three mixing angles $(\theta_{12},\theta_{13},\theta_{23})$ and one CP-violating phase $\delta_{\rm CP}$:
\beq
U = \left ( \ba{ccc} 
c_{12} c_{13} & s_{12} c_{13} & e^{- i \delta_{\rm CP}} s_{13} \\ 
- s_{12} c_{23} - e^{i \delta_{\rm CP}} c_{12} s_{13} s_{23} & c_{12} c_{23}  - e^{i \delta_{\rm CP}} s_{12} s_{13} s_{23}
& c_{13} s_{23} \\ 
s_{12} s_{23} - e^{i \delta_{\rm CP}} c_{12} s_{13} c_{23} & - c_{12} s_{23} - e^{i \delta_{\rm CP}} s_{12} s_{13} c_{23} 
&  c_{13} c_{23} 
\ea \right ), 
\eeq 
and $s_{ij} \equiv \sin \theta_{ij}$, $c_{ij} \equiv \cos \theta_{ij}$. 
As mentioned earlier, the neutrinos have only the left-handed
components, while right-handed neutrinos are absent in this effective
theory. 
The leading NSI corrections to the standard neutrino interactions are summarized by the parameters $\epsilon_X$ in \eref{EFT_lweft}, which are $3 \times 3$ matrices in the lepton flavor space.
In the neutrino literature these are customarily referred to as the NSI.
Apart from the SM-like V-A interactions ($1+\epsilon_L$), right-handed ($\epsilon_R$), scalar ($\epsilon_S$), pseudoscalar ($\epsilon_P$), and tensor ($\hat \epsilon_T$) interactions between leptons and quarks are allowed at the same order of the WEFT expansion.     
The matching between $\epsilon_X$ and the Wilson coefficients of dimension-6 operators in the Warsaw basis~\cite{Grzadkowski:2010es} of the SMEFT  at the renormalization scale $\mu\sim m_W$ is given by~\cite{Cirigliano:2009wk,Cirigliano:2012ab,Aebischer:2017gaw,Gonzalez-Alonso:2017iyc} 
\begin{eqnarray}
\label{eq:EFT_wefttosmeft}
 [\epsilon_L]_{\alpha \beta} & \approx &  
{v^2 \over \Lambda^2 V_{ud}} \left (
V_{ud} [c^{(3)}_{Hl}]_{\alpha\beta} 
+ V_{jd}[c^{(3)}_{Hq}]_{1j} \delta_{\alpha \beta}
- V_{jd} [c^{(3)}_{lq}]_{\alpha\beta 1j} \right) , 
\nnl 
\, [\epsilon_R]_{\alpha\beta}  &\approx & 
{v^2 \over 2 \Lambda^2 V_{ud}} [c_{Hud}]_{11} \delta_{\alpha\beta} ,
\nnl 
\, [\epsilon_S]_{\alpha\beta}
&\approx & -
{v^2 \over 2 \Lambda^2 V_{ud}} \left ( V_{jd} [c^{(1)}_{lequ}]^*_{\beta\alpha j1} +  [c_{ledq}]^{*}_{\beta\alpha 11} \right ),
\nonumber\\
\, [\epsilon_P]_{\alpha\beta}
&\approx  & - {v^2 \over 2 \Lambda^2 V_{ud}} \left ( V_{jd} [c^{(1)}_{lequ}]^*_{\beta\alpha j1} -  [c_{ledq}]^{*}_{\beta\alpha 11} \right ),  
\nnl 
\, [\hat \epsilon_T]_{\alpha\beta}
&\approx  & -   {2 v^2 \over \Lambda^2 V_{ud}} V_{jd}[c^{(3)}_{lequ}]^*_{\beta\alpha j 1}~,
\eea 
where SMEFT operators are defined in the flavor basis where the up-quark Yukawa matrices are diagonal. 
There are three important conclusions from this matching exercise. Firstly, all the $\epsilon_X$ parameters  in \eref{EFT_lweft} arise at $\cO(\Lambda^{-2})$ in the SMEFT, thus a priori they are equally important. 
Secondly, the right-handed interactions are proportional to the unit matrix in the lepton flavor space, up to corrections from dimension-8 and higher SMEFT operators~\cite{Cirigliano:2009wk}.
Indeed, at the dimension-6 level $\epsilon_R$ can originate only from the operator $O_{Hud} = i H^T D_\mu H (\bar u_R \gamma^\mu d_R)$ and its conjugate, which induce the $W$ boson coupling to right-handed quarks. 
Integrating out the $W$ exchange between the quarks and leptons generates $\epsilon_R$ in \eref{EFT_wefttosmeft}.
Since the SM $W$ couplings to leptons are diagonal and flavor universal, 
so is $\epsilon_R$ at $\cO(\Lambda^{-2})$. 
Off-diagonal and flavor non-universal contributions to $\epsilon_R$ can appear only at $\cO(\Lambda^{-4})$, either from the $W$ exchange (if the $W$ couples to right-handed quarks and non-universally to leptons at order $\Lambda^{-2}$), 
or from dimension-8 contact operators such as e.g 
$(\bar L_\alpha H \gamma_\mu L_\beta H)(\bar u_R \gamma^\mu d_R)$, where $L_\alpha = (\nu_L,\ell_L)_\alpha$ are the lepton doublets, and $H$ is the Higgs doublet.  
On the other hand, $\epsilon_{L,S,P,T}$ do contain off-diagonal and non-universal terms already at the dimension-6 level, in general.
Finally, $\epsilon_L$ is approximately a Hermitian matrix in the lepton flavor space, up to corrections suppressed by off-diagonal CKM matrix elements. 
This directly follows from the hermiticity properties of the SMEFT Wilson coefficients: 
$[c^{(3)}_{Hl}]_{\alpha\beta}^* = [c^{(3)}_{Hl}]_{\beta \alpha}$, 
$[c^{(3)}_{Hq}]_{jk}^* = [c^{(3)}_{Hq}]_{kj}$, 
and $[c^{(3)}_{lq}]_{\alpha\beta j k}^* = [c^{(3)}_{lq}]_{\beta \alpha k j}$.

\subsection{Lee-Yang}
\label{sec:Lee-Yang}

At the energy scale characteristic for reactor neutrino experiments the relevant degrees of freedom are not quarks, but rather their bound states such as nucleons and nuclei.
Therefore, it is advantageous to descend one more step in the EFT ladder, into an effective theory of protons and neutrons interacting with charged leptons and neutrinos. 
Matching this EFT to the WEFT Lagrangian in \eref{EFT_lweft} we obtain the so-called Lee-Yang Lagrangian~\cite{Lee:1956qn}:
\bea 
\label{eq:EFT_lleeyang}
\cal L_{\rm LY} & \supset &
-  \frac{V_{ud}}{v^2} \big \{   
g_V \left [ {\bf 1} +    \epsilon_L +  \epsilon_R \right]_{\alpha \beta}  (\bar p  \gamma^\mu n)  (\bar \ell_\alpha  \gamma_\mu P_L \nu_\beta) 
 - g_A \left [ {\bf 1} +    \epsilon_L  -   \epsilon_R \right]_{\alpha \beta}  (\bar p  \gamma^\mu \gamma_5 n)   (\bar \ell_\alpha  \gamma_\mu P_L \nu_\beta) 
\nnl &&
~~~~~+g_S [\epsilon_S]_{\alpha \beta} (\bar p n) (\bar \ell_\alpha P_L  \nu_\beta)  
-  g_P  [\epsilon_P]_{\alpha \beta} (\bar p \gamma_5 n) (\bar \ell_\alpha P_L  \nu_\beta)   
\nnl &&
~~~~~+ {1 \over 2} g_T [\hat \epsilon_T]_{\alpha \beta} (\bar p  \sigma^{\mu \nu} P_L n)   (\bar \ell_\alpha  \sigma_{\mu \nu} P_L \nu_\beta) 
+ \hc  \big \} ,
\eea 
where $p$, $n$ are relativistic proton and neutron fields.
The couplings $g_{V,A,S,P,T}$ are  vector, axial, scalar, pseudoscalar, and tensor charges of the nucleon,  which  can be calculated on the lattice or from symmetry considerations. 
For the vector coupling, one can prove that  $g_V = 1$  up to quadratic corrections in isospin-symmetry breaking \cite{Ademollo:1964sr}. 
For the remaining charges we use the numerical values collected in Table~1 of \cite{Gonzalez-Alonso:2018omy} (which are taken from  Refs.~\cite{Bhattacharya:2016zcn,Gonzalez-Alonso:2013ura}), 
except for $g_A$, which is taken from the fit in Eq.~(84) of~\cite{Gonzalez-Alonso:2018omy}:  
\beq
\label{eq:EFT_gX}
g_A = 1.2728 \pm 0.0017, 
\quad g_S = 1.02 \pm 0.11, \quad  g_P = 349 \pm 9, \quad g_T = 0.987 \pm 0.055,
\eeq 
at $\mu=2$ GeV in the $\overline{MS}$ scheme, a choice of scale and scheme that will apply as well to the $\epsilon_X$ bounds obtained in this work. 
For our purpose, the charges are known with sufficient precision, and we will use their central values ignoring the  errors. 

The processes relevant for neutrino production and detection in reactor experiments are (inverse) beta decays. 
While in these reactions neutrinos and electrons are typically relativistic, the exchanged momenta are much smaller than the masses of nucleons. Therefore, one can  describe the latter using non-relativistic fields $\psi$ in an effective theory expanded in powers of the spatial derivatives $\nabla_k \psi$.
At the leading (zero-derivative) order, 
the Lee-Yang Lagrangian in \eref{EFT_lleeyang} reduces to
\bea 
\label{eq:EFT_NRleeyang}
{\cal L}_{\rm NRLY} & \supset &
-  \frac{V_{ud}}{v^2}   
(\bar \psi_p \psi_n)  \big \{  \left [{\bf 1} +   \epsilon_L +  \epsilon_R \right]_{\alpha \beta}   (\bar \ell_\alpha \gamma^0 P_L \nu_\beta) +g_S [\epsilon_S]_{\alpha \beta}  (\bar \ell_\alpha  P_L \nu_\beta)   \big \} 
\nnl &+ &
 \frac{ V_{ud}}{v^2}     (\bar \psi_p \sigma^k \psi_n)  \big \{
 g_A \left [ {\bf 1 } +   \epsilon_L  -   \epsilon_R \right]_{\alpha \beta}  (\bar \ell_\alpha \gamma^0 \Sigma^k P_L \nu_\beta) 
- g_T [\hat \epsilon_T]_{\alpha \beta}   (\bar \ell_\alpha  \Sigma^k P_L \nu_\beta) 
  \big \}  + \hc , 
\nnl 
\eea 
where $\psi_p$ and $\psi_n$ are non-relativistic fields annihilating protons and neutrons, respectively, and $\Sigma^k = \left ( \ba{cc} 0 & \sigma^k \\ \sigma^k & 0 \ea \right)$ with  $\sigma^k$ being the Pauli matrices.
Note that $\epsilon_P$ does not appear in \eref{EFT_NRleeyang}, hence the pseudoscalar interactions do not affect beta transitions at the leading order. 
Moreover there are only two independent hadronic structures at this order: 
$\bar \psi_p \psi_n$  and $\bar \psi_p \sigma^k \psi_n$, which mediate the Fermi and Gamow-Teller nuclear transitions, respectively.
Continuing the non-relativistic expansion, at the next order one would obtain the interactions with one derivative acting on $\psi$, which  lead to the so-called first-forbidden beta transitions. 

It is worth noting that the {\it same} effective interactions parametrized by $\epsilon_X$ that can be probed in neutrino oscillation experiments also affect the phenomenological extraction of $V_{ud}$ and $g_A$ from nuclear and neutron decays, respectively~\cite{Gonzalez-Alonso:2018omy}. Since the latter quantities are needed to calculate the predicted number of produced and detected neutrino events in oscillation experiments (see~\sref{observables} for more details), these effects have to be taken into account consistently at the chosen order in the EFT expansion. Such a consistent analysis has never been done in previous NSI literature, to the best of our knowledge.
In particular, it can be shown that neutrino oscillation data does not depend on the flavor-diagonal vector EFT couplings $[\epsilon_{L}]_{ee}$ and $\re [\epsilon_{R}]_{ee}$ at any order.
This is so because their {\it direct} effect is completely cancelled by the {\it indirect} effect entering through the phenomenological determination of $V_{ud}$ and $g_A$.
This can be shown at the Lagrangian level, since these nonstandard contributions only appear in the following two combinations~\cite{Gonzalez-Alonso:2018omy}
\bea
\label{eq:EFT_tildevudga}
V_{ud} \left( 1 + [\epsilon_{L}]_{ee} + [\epsilon_{R}]_{ee} \right)~,
\qquad 
g_A \,\re { 1 + [\epsilon_{L}]_{ee} - [\epsilon_{R}]_{ee}\over 1 + [\epsilon_{L}]_{ee} + [\epsilon_{R}]_{ee}}~.
\eea
Precision experiments that provide the numerical values of $V_{ud}$ and $g_A$ in fact measure the above combinations of the SM and NSI parameters,  when interpreted in the EFT context.\footnote{%
In particular, the $g_A$ value in \eref{EFT_gX} is extracted from experiment and does include such effects of $[\epsilon_{L,R}]_{ee}$.}
For this reason the effects of $[\epsilon_{L}]_{ee}$ and $ [\epsilon_{R}]_{ee}$ are completely absorbed into the phenomenological values of $V_{ud}$ and $g_A$, and thus cannot be accessed in neutrino oscillation experiments. 
These observations invalidate  bounds on $[\epsilon_{L,R}]_{ee}$ obtained from oscillation experiments in previous literature (see e.g. Table 4 in Ref.~\cite{Farzan:2017xzy}). 
Let us note that the $[\epsilon_{L,R}]_{ee}$ coefficients can be probed in precision beta decay measurements, through a (first-row) CKM unitarity test and through the comparison of lattice and experimental values of $g_A$. The resulting bounds are below the permil and percent level, respectively~\cite{Gonzalez-Alonso:2018omy}.

In the following we will use the Lagrangian in \eref{EFT_NRleeyang} to calculate amplitudes of beta decay processes relevant for reactor neutrino oscillations. 
We will treat $\epsilon_X$ as small parameters of order $\Lambda^{-2}$, as  derived from the matching to the SMEFT, and we will ignore any contributions to observables that are $\cO(\Lambda^{-4})$ or smaller.

\section{Oscillations in EFT}
\label{sec:oscillations}
\setcounter{equation}{0}

In this section we review the theory of neutrino oscillations in the presence of NSI. 
We focus on providing a systematic EFT description of new physics effects in neutrino production and detection in short-baseline reactor experiments.  
We neglect matter effects in neutrino propagation, which would be relevant for long-baseline experiments.

Consider an antineutrino produced with energy $E_\nu$ in the process 
$X^P \to \ell_\alpha^- \bar \nu Y^P$ and detected in the process  $\bar\nu  X^D  \to \ell_\alpha^+ Y^D$,
where $\ell_\alpha$ is a charged lepton: electron, muon, or tau. 
Given a neutrino produced in association with $\ell_\alpha^-$, the {\em survival probability} is defined as the probability of it being detected at the distance $L$ from the source in association with  $\ell_\alpha^+$ of the same  lepton flavor. 
Quite generally, the survival probability is given by the formula~\cite{Giunti:2007ry}:
\beq
\label{eq:OSC_master}
P_{\bar \nu_\alpha \to \bar \nu_\alpha}(L,E_\nu)  =   
 \sum_{JK} C_{JK}^{\alpha}  \exp \left (-i {\Delta m_{JK} ^2 L \over 2 E_\nu } \right ),
 \qquad 
 C_{JK}^{\alpha} \equiv { (\int A_{\alpha J}^P  A_{\alpha K}^{P \, *} )  (\int A_{J \alpha}^D  A_{K \alpha }^{D\, *}) \over  \left (\sum_{I} \int| A_{\alpha I}^P |^2 \right ) \left (\sum_{I'} \int |A_{I' \alpha }^D |^2 \right )}, 
 \eeq 
 where the indices $J,K,\dots$ label neutrino mass eigenstates $\nu_J$,  $\Delta m_{JK}^2  \equiv  m_{\nu_J} ^2 - m_{\nu_K}^2$, and $A^{P}_{\alpha J}$ and $A^{D}_{J\alpha }$ denote the amplitudes for the production and detection of $\nu_J$:   
\beq
\label{eq:OSC_apd}
A_{\alpha J}^P  \equiv  \cM(X^P \to \ell_\alpha^- \bar \nu_J Y^P), 
\qquad 
A_{J \alpha}^D  \equiv   \cM(\bar \nu_J X^D \to \ell_\alpha^+ Y^D ).
\eeq  
In neutrino oscillation experiments, polarization of particles involved in production and detection is not measured, therefore summation over spins (and any other internal indices) is implicit in each bracket in \eref{OSC_master}. Likewise, there is an integration over all kinematic variables (except the neutrino energy), as indicated by 
$\int$ in \eref{OSC_master}.

We now derive a general expression for the coefficients 
$C_{JK}^{\alpha}$ as a function of the Wilson coefficients $\epsilon_X$ in the WEFT Lagrangian \eref{EFT_lweft}. 
The amplitudes in \eref{OSC_apd} can be decomposed as 
\beq
A_{\alpha J}^P = U_{\alpha J} M_L^P +  \sum_{X = L,R,S,P,T} [\epsilon_X U]_{\alpha J} M_X^P, 
\qquad 
A_{J \alpha }^D   =   U^\dagger_{J \alpha} M_L^D +  \sum_{X = L,R,S,P,T}  [U^\dagger \epsilon_X^\dagger]_{J \alpha} M_X^D. 
\eeq  
Here $M_X^P$ and $M_X^D$  are independent of the mass index of the emitted/absorbed antineutrino, up to totally negligible corrections due to the  neutrino masses.
 Then, keeping only the linear effects in $\epsilon_X$, we can approximate  
\bea 
\label{eq:OSC_cg} 
C_{JK}^{\alpha} & = &  U_{\alpha J}  U^\dagger_{K \alpha}   U_{\alpha K}  U^\dagger_{J \alpha}
\nnl & + &  U_{\alpha K}  U^\dagger_{J \alpha}  \sum_{X =  L,R,S,P,T} \sum_{\gamma \neq \alpha }\left \{ 
p_X  [\epsilon_X]_{\alpha \gamma} U_{\gamma J } U^\dagger_{K \alpha} + 
p_X^* U_{\alpha J} U^\dagger_{K \gamma} [\epsilon_X^\dagger]_{\gamma \alpha}     \right \}
\nnl & + &
  U_{\alpha J}  U^\dagger_{K \alpha}    \sum_{X =  L,R,S,P,T}  \sum_{\gamma \neq \alpha } \left \{d_X^* [\epsilon_X]_{\alpha \gamma} U_{\gamma K} U^\dagger_{J \alpha}  + d_X U_{\alpha K} U^\dagger_{J \gamma} [\epsilon_X^\dagger]_{\gamma \alpha}   \right \} 
  + \cO(\epsilon_X^2)~,
\eea
where
\bea 
p_X \equiv  {\int M_X^P M_L^{P \, *} \over \int |M_L^P|^2 }~, \qquad 
d_X \equiv {\int M_X^D M_L^{D\, *} \over \int |M_L^D|^2 }~. 
\eea 
The first line in~\eref{OSC_cg} encapsulates the standard oscillations  in the absence of BSM effects other than the neutrino masses.     
The second and third lines in~\eref{OSC_cg} describe corrections to the survival probability due to NSI affecting, respectively, the neutrino production and detection processes.   
The coefficients $p_X$ and $d_X$ depend on the processes in which neutrinos are produced and detected, and in general they may be functions of the neutrino energy. 
Note that the diagonal elements $\epsilon_X$ do not enter \eref{OSC_cg}; in fact they cancel out between the numerator and denominator of \eref{OSC_master}.
Therefore, only the off-diagonal (in the charged-lepton flavor basis) Wilson coefficients of the effective Lagrangian \eref{EFT_lweft} affect the survival probability at the leading order.
Recall that if the WEFT Lagrangian is derived from the underlying SMEFT (that is, if new physics is heavier than $m_W$ and respects the full SM local symmetry) then $\epsilon_R$ is a diagonal matrix, leading to the  conclusion  that the charged currents involving right-handed quarks do not affect neutrino oscillations at $\cO(\Lambda^{-2})$.   

Specializing to reactor experiments such as Daya Bay  and RENO, neutrinos are detected via inverse beta decay on water (practically, proton) targets, with a positron and a neutron in the final state:  $\bar \nu p \to n e^+$. 
Calculating the amplitude for this process starting from the non-relativistic effective Lagrangian in  \eref{EFT_NRleeyang} we find the following detection coefficients
\beq
\label{eq:OSC_dx}
d_L \equiv 1,
\quad 
d_R = -  {3 g_A^2 - 1  \over 3 g_A^2 + 1}, 
\quad 
d_S = - {g_S \over 3 g_A^2 + 1} {m_e \over E_\nu - \Delta}, 
\quad 
d_T =  {3 g_A g_T  \over 3 g_A^2 + 1} {m_e \over E_\nu - \Delta}, 
\quad
d_P = 0, 
\eeq 
where $\Delta \equiv m_n - m_p \approx 1.29$~MeV and $m_e \approx 0.511$~MeV is the positron mass. 
The same result is obtained starting from the relativistic \eref{EFT_lleeyang} in the limit  where the proton recoil is neglected.  
That calculation reveals that the contribution proportional to $\epsilon_P$ is suppressed by the small  factor $g_P m_e/m_p \sim 0.1$ in spite of the large value of the pseudoscalar charge $g_P$. 
In the following we neglect these subleading pseudoscalar contributions.
Note that $d_S$ and $d_T$ depend on the neutrino energy, which will be an important handle for constraining the scalar and tensor Wilson coefficients in reactor experiments. 
The factor in the amplitude proportional to ${m_e \over E_\nu - \Delta} \approx {m_e \over E_e}$ goes under the name of the {\em Fierz interference term}~\cite{Fierz1937}, and is due to the lepton-chirality flip in the corresponding Lagrangian terms in  \eref{EFT_NRleeyang}.

While non-standard effects on the detection side are calculable to a good accuracy, the production side is far more involved. 
There are hundreds of different beta decay processes contributing to the antineutrino flux in the reactor~\cite{Mueller:2011nm,Huber:2011wv}, and the NSI effects on their amplitudes may be subject to relatively large uncertainties.
To tackle that problem, we have to resort to certain crude approximations. 
First, we assume that all beta decays contributing to the reactor antineutrino flux above the detection threshold $E_\nu = 1.8$~MeV  are of the  Gamow-Teller type. 
With that assumption, 
the production coefficients are given by 
\beq\label{eq:prodCoef}
p_L \equiv  1, \qquad p_R = -1, \qquad  p_S \approx 0, \qquad p_P \approx 0, \qquad p_T= -{g_T \over g_A} {m_e \over f_T(E_\nu) } . 
\eeq 
As before, the pseudoscalar interactions can be neglected at the leading order. 
In  addition the scalar ones do not contribute to Gamow-Teller transitions. The form factor in the tensor coefficient is  given by
\bea 
\label{eq:OSC_ftenu}
 f_T(E_\nu) 
 &=&
 {
 \sum_{i=1}^n  w_i  (\Delta_i  - E_\nu) \sqrt{ (\Delta_i  - E_\nu - m_e) (\Delta_i  - E_\nu + m_e) }
  \over 
 \sum_{i=1}^n  w_i   \sqrt{ (\Delta_i  - E_\nu - m_e) (\Delta_i  - E_\nu + m_e)}
 }\nnl
 &\approx&
 {
 \int_{E_\nu+m_e}^\infty d\Delta \,W(\Delta)  (\Delta  - E_\nu) \sqrt{ (\Delta  - E_\nu - m_e) (\Delta  - E_\nu + m_e) }
  \over 
 \int_{E_\nu+m_e}^\infty d\Delta \,W(\Delta)   \sqrt{ (\Delta  - E_\nu - m_e) (\Delta  - E_\nu + m_e)}
 }~.
\eea
The sum in the first line goes over all $\beta$ decays resulting from nuclear fission processes in the reactor, with appropriate weight factors $w_i$ determined by the fission yield. 
Furthermore, $\Delta_i$ are the mass differences of the initial and final state nuclei participating in the beta processes.
As shown in the second line of~\eref{OSC_ftenu}, rather than using a detailed reactor model with all distinct processes explicitly included, to calculate $f_T(E_\nu)$ we  
replace the sums by integrals over endpoint energies. We use a gaussian distribution for $W(\Delta)$~\cite{King:1958zz} peaked at $1.7$~MeV and with $\sigma=2.5$~MeV, which approximates well the phenomenological distribution (see e.g. Refs.~\cite{Davis:1979gg,Vogel:2007du}).

In reality, only about 70\% of beta transitions in reactors  are of the  Gamow-Teller type~\cite{Hayes:2016qnu}. 
Most of the remaining ones are the first-forbidden transitions, whose neutrino spectrum has considerable uncertainties even in the SM limit, and whose dependence on non-standard interactions is poorly known (see Ref.~\cite{Glick-Magid:2016rsv} for recent work in this direction). 
These are expected to give a non-negligible contribution to the reactor antineutrino flux, especially for $E_\nu$ far above the detection threshold~\cite{Hayen:2018uyg}.  
In particular, the first-forbidden decays may reintroduce some sensitivity to the pseudoscalar interactions. 
In this paper we ignore this complication, however we will check the robustness of our results by testing how much they rely on the events at the high end of the reactor antineutrino spectrum.

In the SMEFT approach one assumes no new degrees of freedom beyond those of the SM, therefore the sum in \eref{OSC_master} goes over the 3 neutrino states, and the oscillation  probability in general depends on the two independent mass squared differences $\Delta m_{21}^2$ and $\Delta m_{31}^2$.  
However, in short-baseline neutrino experiments one can typically neglect the effects proportional to $\Delta m_{21}^2 L/E$;   
in particular, this is a good approximation in  Daya Bay and RENO. 
In such a case \eref{OSC_master} simplifies to 
\bea
\label{eq:OSC_2flavor} 
P_{\bar \nu_\alpha \to \bar \nu_\alpha}(L,E_\nu)&  \approx & 
C_{11}^{\alpha} +  C_{22}^{\alpha}   + C_{33}^{\alpha} 
+ 2 \,\re \left ( C_{12}^{\alpha} +  C_{13}^{\alpha}   + C_{23}^{\alpha}  \right )
- 4  \,\re \left (C_{13}^{\alpha} + C_{23}^{\alpha}  \right ) \sin^2 \left (\Delta m_{31}^2 L \over 4 E_\nu \right )
\nnl & - &  
   2  \,\im \left ( C_{13}^{\alpha} +  C_{23}^{\alpha}  \right ) \sin \left (\Delta m_{31}^2 L \over 2 E_\nu \right )
+ \cO\left (\Delta m_{21}^2 L \over E_\nu \right ). 
  \eea 
For the reactor experiments the relevant observable is the electron antineutrino survival probability. 
Taking $\alpha  = e$, and plugging in the expression of $C^{e}_{JK}$ in \eref{OSC_cg}  we obtain 
\bea
\label{eq:OSC_dxpx}
P_{\bar \nu_e \to \bar \nu_e}(L,E_\nu)  & = & 1 -  \sin^2 \left (\Delta m_{31}^2 L \over 4 E_\nu \right )
\sin^2 \left ( 2 \theta_{13} +  \sum_{X=L,S,T} (d_X + p_X) \re [X]  \right ) 
\nnl & - & 
  \sin \left (\Delta m_{31}^2 L \over 2 E_\nu \right )  \sin(2 \theta_{13}) \sum_{X=L,S,T} (d_X -  p_X) \im [X]    + \cO(\eps_X^2) , 
\eea  
where we defined the following combinations of the PMNS and NSI parameters:
\bea  
\label{eq:OSC_ST}
\, [L] &\equiv&   e^{i \delta_{\rm CP}} 
\left (s_{23} [\epsilon_L]_{e \mu} + c_{23} [\epsilon_L]_{e \tau}  \right )~,\nnl
\, [S] &\equiv&   e^{i \delta_{\rm CP}} 
\left (s_{23} [\epsilon_S]_{e \mu} + c_{23} [\epsilon_S]_{e \tau}  \right )~,\nnl
\, [T] &\equiv&   e^{i \delta_{\rm CP}} 
\left (s_{23} [\hat \epsilon_T]_{e \mu} + c_{23} [\hat \epsilon_T]_{e \tau}  \right )~.
\eea
Using the detection and production coefficients in Eqs. (\ref{eq:OSC_dx}) and (\ref{eq:prodCoef}), the survival probability takes the form 
\bea
\label{eq:OSC_generaltemplate}
P_{\bar \nu_e \to \bar \nu_e}(L,E_\nu)
  & =  & 
1 
- \sin^2 \left (\Delta m_{31}^2 L \over 4 E_\nu \right ) \sin^2 \left ( 2 \tilde \theta_{13} -  \alpha_D  {m_e \over E_\nu - \Delta}  -   \alpha_P {m_e \over f_T(E_\nu)}  \right) 
\nnl 
&+ & \sin \left (\Delta m_{31}^2 L \over 2 E_\nu \right )   \sin (2 \tilde \theta_{13})  \left ( \beta_D {m_e \over E_\nu - \Delta} - \beta_P {m_e \over f_T(E_\nu)}  \right ) + \cO(\epsilon_X^2) , 
\eea 
where
\bea
\label{eq:OSC_generaldictionary}
&\tilde \theta_{13}  =  \theta_{13} +  \re \left  [L  \right ],~~~~~~~~~~~~~~~~~~~~~~~~&  
 \nnl 
 &\alpha_D  =   
 {g_S \over 3 g_A^2 + 1}  \re \left [ S \right ]
 -  {3 g_A g_T \over 3 g_A^2 + 1}  \re \left [ T \right ] ,
 \qquad
 &\alpha_P =     {g_T \over g_A}  \re \left [ T \right ] ,
 \nnl  
 &\beta_D  = 
 {g_S \over 3 g_A^2 + 1}   \im \left [ S \right ]
-  {3 g_A g_T \over 3 g_A^2 + 1}   \im \left [ T \right ] , 
 \qquad
 &\beta_P =    {g_T \over g_A}   \im \left [ T \right ] , 
\eea 
The oscillation formula in~\eref{OSC_generaltemplate} is valid away from $\Delta m_{31}^2 L/E_\nu \approx 0$, when the Wilson coefficients $\epsilon_X$ obey the SMEFT scaling: $\epsilon_X^2 \sim \cO(\Lambda^{-4}) \ll \epsilon_X \sim \cO(\Lambda^{-2})$, and for $d_X [X], p_X [X]  \ll \theta_{13}$.
In its derivation we assumed that the off-diagonal elements of $\epsilon_R$ vanish, which is true up to $\cO(\Lambda^{-4})$ corrections  when the SMEFT is a valid effective theory in some energy regime above $m_W$ (see \eref{OSC_generaltemplate2} for a more general formula).  
The expression for $P_{\nu_e \to \nu_e}$ would be analogous with the reversed sign of the second line of \eref{OSC_generaltemplate}.
Our formula agrees with the survival probability written down in Ref.~\cite{Kopp:2007ne}
after expressing their effective couplings $\epsilon^{s,d}$ by the Wilson coefficients of the WEFT Lagrangian, see \aref{NSInu}.

There are several important conclusions one can draw from \eref{OSC_generaltemplate}:
\begin{itemize}
\item As mentioned before, the neutrino survival probability at the leading order depends only on off-diagonal Wilson coefficients $\epsilon_X$. 
We remark that the total number of produced and detected events (rather than the survival probability) is in principle sensitive to the diagonal scalar and tensor $[\epsilon_{S,T}]_{ee}$, which we discuss in more detail in~\sref{observables}. However, this caveat has no practical consequences due to the very stringent model-independent constraints on these coefficients from nuclear and meson decays~\cite{Gonzalez-Alonso:2018omy}.
As discussed below \eref{EFT_tildevudga}, the effects of  $[\epsilon_L]_{ee}$ and $[\epsilon_R]_{ee}$ are completely absorbed into the phenomenological values of the CKM element $V_{ud}$ and the axial charge of the nucleon $g_A$, and are unobservable in neutrino oscillation experiments. 

\item 
The sensitivity of reactor experiments  to pseudoscalar NSI  ($\epsilon_P\neq0$) vanishes in the zero-recoil limit of beta decays, and when first-forbidden transitions in the reactor are neglected.

\item  At the leading order, reactor experiments alone are {\em not} sensitive to off-diagonal NSI of the V-A type ($[\epsilon_L]_{e \alpha} \neq 0$).
The reason is that, as evident in \eref{OSC_generaldictionary}, their effects can be fully absorbed into a redefinition of the PMNS mixing angle $\theta_{13}$ into the effective mixing angle $\tilde \theta_{13}$.\footnote{%
This issue is well-known in electroweak precision measurements (see e.g. \cite{Wells:2005vk}), 
where some non-standard effects may be absorbed into a redefinition of the  SM parameters.
For example, the $\cO(10^{-7})$ measurement of the Fermi constant from the muon lifetime does not constrain new physics at this precision level, as the non-standard corrections can be absorbed into a redefiniton of an a-priori unknown electroweak parameter - the Higgs vaccuum expectation value. 
For analogous effects in CKM physics, see Ref.~\cite{Descotes-Genon:2018foz}.} 
That redefinition can in fact be performed including also quadratic corrections in $\epsilon_L$~\cite{Ohlsson:2008gx}.
Since $\theta_{13}$ is an unknown parameter, which in the standard context was actually {\em measured} by  Daya Bay, RENO, and Double Chooz, these experiments cannot separate the effect of the PMNS mixing parametrized by $\theta_{13}$ from the new physics corrections contained in $[\epsilon_L]_{e \alpha}$.
To that end, it is necessary to measure another observable that is sensitive to a {\em different} combination of $\theta_{13}$ and $\epsilon_L$ than the one defined by $\tilde \theta_{13}$. 
This conclusion continues to hold when  subleading terms in $\Delta m^2_{21}$ are taken into account in the survival probability. 

\item 
On the other hand, reactor experiments are sensitive to scalar and tensor charged-current interactions between leptons and quarks.
The survival probability depends on the real and imaginary parts of the $[S]$ and $[T]$ combinations defined in~\eref{OSC_ST}. 
Two handles allow us to  explore that dependence in practice. 
One, in the presence of CP violation (due to $\delta_{\rm CP}$ in the PMNS matrix or imaginary components of $\epsilon_{S,T}$), the survival probability acquires a different oscillatory dependence on $L/E_\nu$ than in the standard case. 
Secondly, the knowledge of the dependence of the survival  probability on neutrino energy $E_\nu$ allows one to disentangle CP-conserving effects of scalar and tensor interactions from each other, and from the (energy-independent) effective mixing angle $\tilde \theta_{13}$.  

\item The survival probability in \eref{OSC_generaltemplate} manifestly satisfies $0 < P(\bar \nu_e \to \bar \nu_e) \leq 1$ in its regime of validity specified below \eref{OSC_generaldictionary}. 
Naively, for $\Delta m_{31}^2 L/E_\nu \ll 1$ one could obtain $P(\bar \nu_e \to \bar \nu_e) > 1$ or $P(\nu_e \to  \nu_e) > 1$ (depending on the sign and magnitude of $\beta_X$) due to the contribution in the second line in \eref{OSC_generaltemplate}. 
In this regime, however, one can show that the $\cO(\epsilon_X^2)$ contributions cannot be neglected; including the full non-linear $\epsilon_X$ dependence in \eref{OSC_cg} one recovers $P(\bar \nu_e \to \bar \nu_e) \leq 1$ independently of the magnitude of $\beta_X$. 
Note that the Daya Bay and RENO experiments are designed such that $\Delta m_{31}^2 L /E_\nu \sim 1$ for typical $E_\nu$, 
therefore this caveat has no practical consequences for our analysis. 
Note also that there are no non-oscillatory terms in \eref{OSC_generaltemplate}, therefore the so-called zero-distance effects~\cite{Langacker:1988ur,Kopp:2007ne} sometimes discussed in the NSI literature are absent in our approach. 
This is reassuring, as zero-distance effects at the linear level in $\epsilon_X$ would also lead to $P(\bar \nu_e \to \bar \nu_e) > 1$ for some parameter choices.  
\item
The last term in the survival probability in \eref{OSC_generaltemplate} is proportional to $\sin(\Delta m^2_{31}L/(2E_\nu))$, which clearly depends on the choice of mass ordering. Throughout this analysis we assume $\Delta m^2_{31}>0$. 
Choosing the inverted mass ordering would result in the opposite signs for  the best fit values of  $\im[S]$ and $\im[T]$ compared to that determined in the next section. 
  \end{itemize}

\begin{figure}[t]
\centering
\includegraphics[width=1\textwidth]{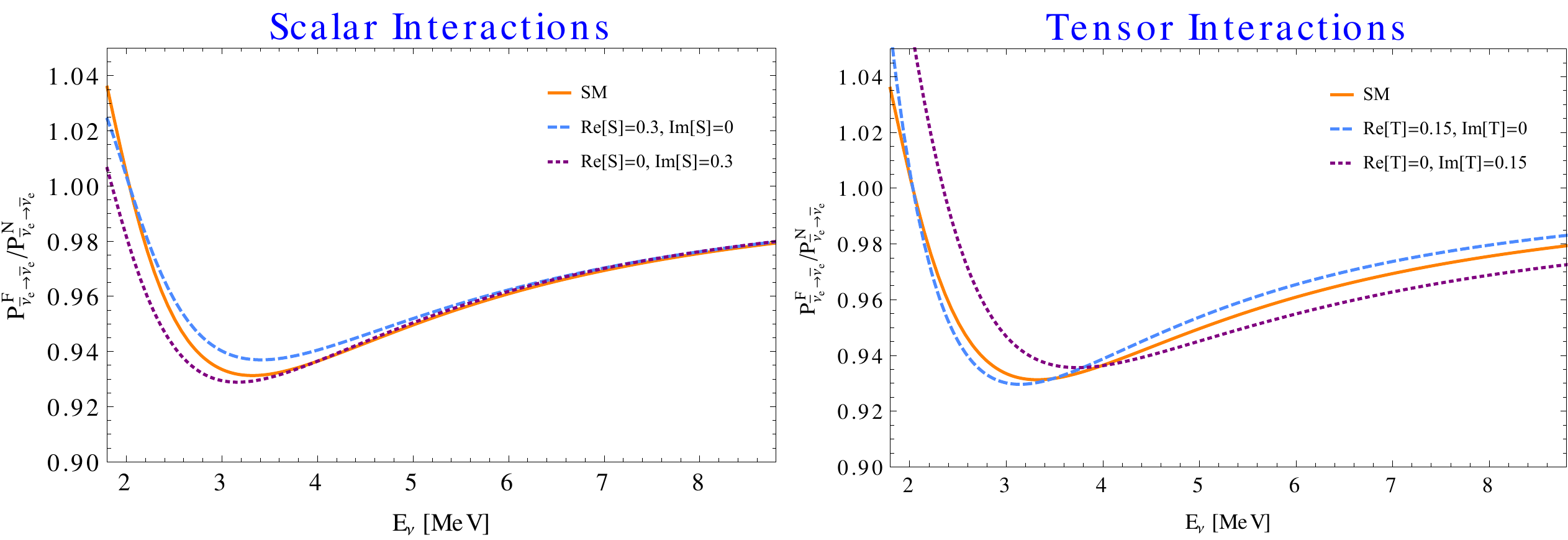}
\caption{\label{fig:probability}
The ratio of the survival probabilities at the far and near sites as a function of the neutrino energy and at distances $L=1500\,\rm{m}$ and $L=500\,\rm{m}$.
The solid orange curves are for the SM best fit value, while the dashed blue and dotted purple curves show the NSI effects. 
}
\end{figure}

For the sake of illustration, in Fig.~\ref{fig:probability} we show the impact of NSI  on the ratio of the far to near survival probabilities as a function of the neutrino energy assuming a far and near detector distances of $L=1500\,\rm{m}$ and $L=500\,\rm{m}$, respectively. 
The left and right panels show the effects of scalar and tensor interactions. 
To generate these  plots we used the best fit values of the oscillation parameters $\Delta m^2_{31}$ and $\theta_{13}$ from the recent global fit of the neutrino oscillation data in Ref.~\cite{Esteban:2018azc}. 
The orange curves are the SM probability without any NSI effects. 
The dashed blue and dotted purple curves show the effect of the real and imaginary parts of the combination of parameters $[S]$ and $[T]$ defined in \eref{OSC_ST}. 
We see that in both cases the effect of both $\re[X]$ and $\im[X]$ is not only to shift the survival probability, but also distort its $E_\nu$ spectrum  due to the different energy dependence of the scalar and tensor interactions compared to the SM one.  
Note that, as can be seen from the right panel of Fig.~\ref{fig:probability}, reactor neutrino oscillations are more sensitive to tensor interactions. 
This is because they interfere with the SM axial interactions, which typically give larger contributions than the SM vector ones  in reactor transitions and inverse beta decay. 
In the scalar case (the left panel) the far to near probability is similarly sensitive to the change in the real and imaginary parts. This is because the contributions of the real and imaginary parts consist of single terms (coming only from the detection side) which are of the same order and have similar effects on the probability. 
This is not the case for the tensor case (right panel), for which the real and imaginary parts contributions appear as the sum of two terms: those two terms have opposite sign for the real part but the  same sign for the imaginary part.
For this reason the survival probability is more sensitive to the imaginary part of $[T]$.
These comments are illustrative, as they they are valid for a fixed value of $\tilde\theta_{13}$, and they can change in a complete analysis where the latter is also a floating parameter, as we will see in~\sref{results}.

\section{Constraints on EFT parameters from oscillations}
\label{sec:fit}
\setcounter{equation}{0}

\subsection{Observables and NSI sensitivity}
\label{sec:observables}
Typical reactor experiments detect antineutrinos via the inverse beta decay (IBD) process  $\bar \nu p \to e^+ n$. They can measure not only the number of events but also the antineutrino energy. 
In the EFT framework the number of detected IBD events with an antineutrino-energy $E_\nu$ at a distance $d$ is given by
\beq
d N^{\rm EFT}(d,E_\nu)
= 
\rho(d,E_\nu) P_{\bar{\nu}_e\to\bar{\nu}_e}(d,E_\nu)\,d E_\nu\, , \qquad \rho \equiv {d N_\nu^{\rm no-osc} \over d E_\nu} \, ,
\eeq 
where $P$ is the survival probability given in Eq.~(\ref{eq:OSC_generaltemplate}), and $d N_\nu^{\rm no-osc}$ is the differential number of IBD events that would be  detected in the  absence of oscillations.\footnote{
We note that zero-distance effects are included in $P$ and not in $d N_\nu^{\rm no-osc}$, since they are due to a mismatch between the source and detector neutrino flavor eigenstates. We recall that these effects do not appear at $\cO (\Lambda^{-2})$.
}

Note that the latter depends on the distance in a purely geometric form ($\sim 1/d^2$). It also depends on the nonstandard EFT coefficients through the nuclear decays widths and the IBD detection cross section. However, the situation is the opposite as in the survival probability, since $N_\nu^{\rm no-osc}$ does depend linearly (order $\Lambda^{-2}$) on flavor-diagonal coefficients $[\epsilon_X]_{ee}$, but not on flavor-nondiagonal ones. Moreover, since there is no dependence at any order in the vector EFT couplings $[\epsilon_{L,R}]_{ee}$ (see discussion at the end of~\sref{Lee-Yang}), the only linear BSM corrections in $N_\nu^{\rm no-osc}$ comes from flavor-diagonal scalar and tensor coefficients, $[\epsilon_S]_{ee}$ and $[\hat \epsilon_T]_{ee}$. We have taken into account once again that the pseudo-scalar contribution vanishes in the non-relativistic limit and we neglect the contribution of forbidden decays.

To reduce systematic errors, reactor experiments use near and far detectors at different distances from the reactor sources. 
The ratio of the number of IBD events in the energy bin around $E_\nu = \bar E_\nu^i$ in two such detectors is given by 
\beq
\label{eq:FIT_Tijk}
T_i^{j/k} \equiv { N^{\rm EFT}_i (d_j) \over N^{\rm EFT}_i(d_k)} 
= 
{\int_{\bar E_\nu^i}^{\bar E_\nu^i + \Delta} d E_\nu \rho(d_j,E_\nu) P(d_j,E_\nu) 
\over 
\int_{\bar E_\nu^i}^{\bar E_\nu^i + \Delta} d E_\nu \rho(d_k,E_\nu) P(d_k,E_\nu)} 
\approx 
\left( {d_k \over d_j}\right)^2 
{P_{\bar \nu_e \to \bar \nu_e}(d_j,\bar E_\nu^i) \over 
P_{\bar \nu_e \to \bar \nu_e}(d_k,\bar E_\nu^i) }~,
\eeq 
where the last approximation is valid for small enough energy bins $\Delta$. 
In that case, the $\rho$ contributions to the numerator and denominator cancel except for the geometric $d$-dependence (flux), and only the EFT corrections entering via the survival probability $P_{\bar \nu_e \to \bar \nu_e}$ have an effect on the ratio $T_i^{j/k}$. 
Note however that for large energy bins, EFT corrections to $\rho$ with an energy dependence different to the SM one (as is the case for scalar and tensor interactions) do not cancel. 
Obviously, this is also the case for the ratio of the inclusive number of events in two detectors, since one has to integrate over all energies. 
Finally, we note that there is an additional linear effect of flavor-diagonal interactions in the total number of events, which (i) cancels in the far/near ratios and (ii) suffers the large uncertainty of the total reactor flux. 

This introduces a dependence of the inclusive detector rates (and their far/near ratios) on the diagonal coefficients $[\epsilon_S]_{ee}$ and $[\hat \epsilon_T]_{ee}$. 
However, given the relatively large (at least percent level) uncertainties in reactor nuclear processes, and the strong (per-mille level or better) constraints from \lq\lq cleaner" beta decays and other precision experiments on these diagonal coefficients~\cite{Gonzalez-Alonso:2018omy}, 
we will simply ignore this effect in our analysis. 
In fact, for this same reason, such diagonal EFT coefficients cannot explain the observed deficit of detected reactor antineutrino fluxes relative to the SM predictions~\cite{Mueller:2011nm,Huber:2011wv}, which is often referred to as the reactor antineutrino anomaly~\cite{Mention:2011rk}. 

\subsection{Setup and analysis}

In our numerical analysis we use the results from the Daya Bay~\cite{Adey:2018zwh} and RENO~\cite{Bak:2018ydk} experiments with 1958 days and 2200 days of data taking, respectively. The Daya Bay experiment has 4 near detectors located at Experimental Halls 1 and 2 (EH1 and EH2) and 4 far detectors located at Experimental Hall 3 (EH3). The weighted distances from the reactor cores are respectively $516$~m, $555$~m and $1571$~m. The RENO experiment has one near and one far detector located at $367$~m and $1440$~m, respectively.

First we define the following $\chi^2$ function that only uses spectral information
\beq 
\label{eq:chisq}
\chi^2_{\rm spectral}
=
\sum_{i=1}^{N_{\rm{bins}}} \left( \frac{R_{i,obs}^{F/N}-R_{i,th}^{F/N}}{ \delta R_i^{F/N}}\right)^2
+\sum_{d}\Big(\frac{b^d}{\sigma^d_{\rm{bkg}}}\Big)^2+\sum_{r}\Big(\frac{f_r}{\sigma^{r}_{\rm{flux}}}\Big)^2+\Big(\frac{\epsilon}{\sigma_{eff}}\Big)^2\,   
\eeq 
where $R_i^{F/N}=N_{i}^F/N_{i}^N$ is the ratio of far to near IBD events in the $i$-th bin of energy. 
For Daya Bay, since there are two sets of near detectors at different distances, one defines $N_i^N\equiv \omega_{\rm{EH1}}N_i^{\rm{EH1}}+\omega_{\rm{EH2}}N_i^{\rm{EH2}}$, 
where $\omega_{\rm{EH1}}=0.05545$ and $\omega_{\rm{EH2}}=0.2057$ are the weights that sample the different fluxes of the different reactors in equal proportions to the two near experimental halls~\cite{An:2016ses}. 
The statistical uncertainty $\delta R_i^{F/N}$ is given by 
\bea
\delta R_i^{F/N}=\frac{N_{i,obs}^{F}}{N_{i,obs}^N}\sqrt{\frac{N_{i,obs}^{F}+N_{i,bkg}^{F}}{(N_{i,obs}^{F})^2}+\frac{N_{i,obs}^{N}+N_{i,bkg}^{N}}{(N_{i,obs}^{N})^2}}\,,
\eea 
where $N_{i,bkg}$ is the background expected in each energy bin. 
The systematic uncertainties of the background, reactor flux, and efficiency are taken into account by the pull parameters $b^d$, $f_r$, and $\epsilon$, respectively, which we take from the original Daya Bay and RENO publication~\cite{An:2016ses,Bak:2018ydk}. The $d$ and $r$ indices refer to the different detectors and reactors. 
Finally, we construct $\chi^2_{\rm spectral}$ separately for RENO and Daya Bay, 
and combine the two in our analysis.  

In addition to the spectral information in $\chi^2_{\rm spectral}$, we also take into account the ratio of the total IBD rate measured in the near and far detectors of Daya Bay and RENO, following closely the method described in Ref.~\cite{Esmaili:2013yea}. For our analysis we simply sum the two likelihoods: $\chi^2 = \chi^2_{\rm spectral} + \chi^2_{\rm rate}$.

\subsection{Results}
\label{sec:results}

\begin{figure}[!tb]
\begin{center}
\includegraphics[width=1\textwidth]{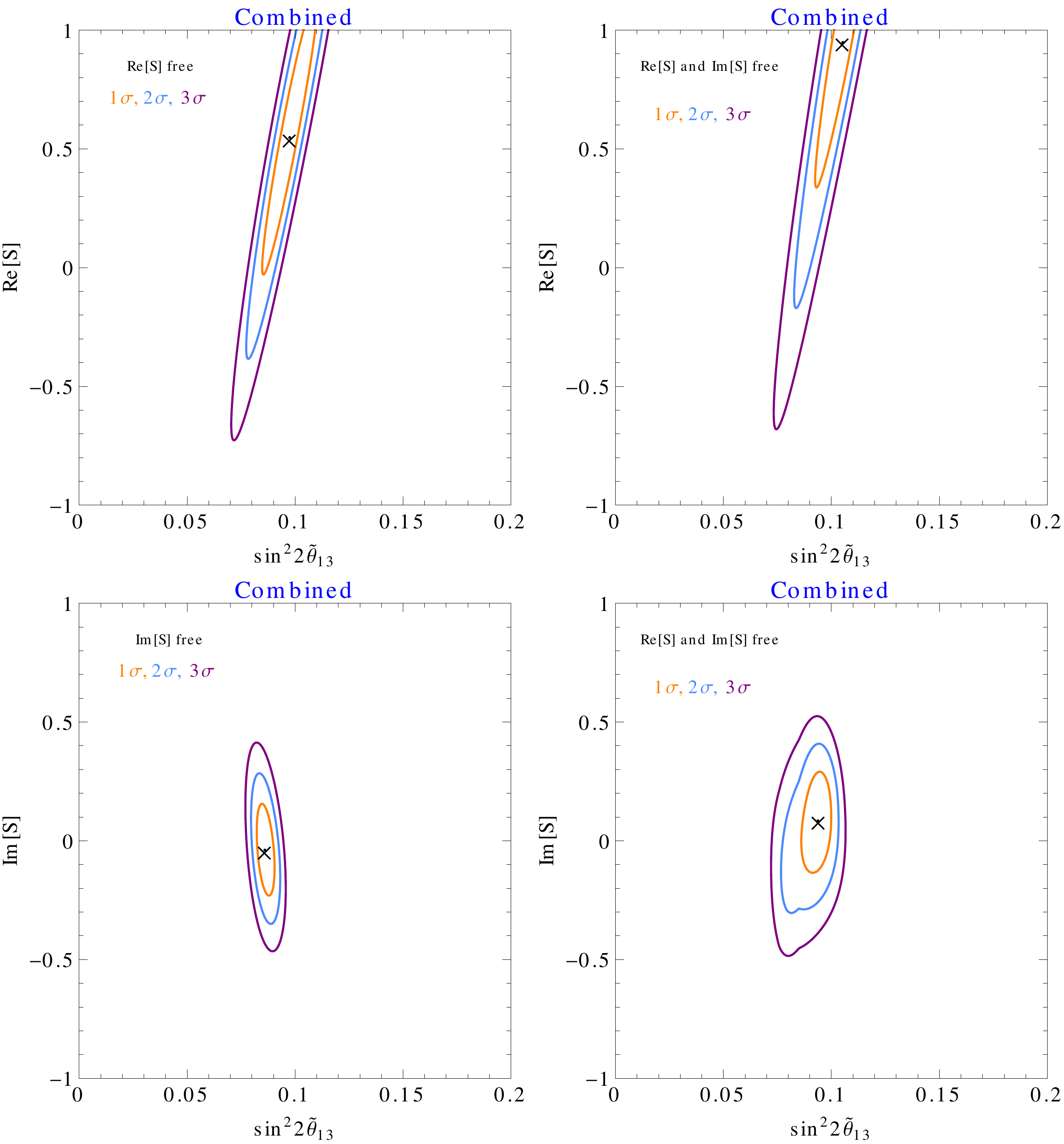}
\end{center}
\caption{
\label{fig:SALL}
\footnotesize{Allowed regions in the $(\sin^22\tilde\theta_{13}-\re[S])$ (first row) and $(\sin^22\tilde\theta_{13}-\im[S])$ plane (second row) for the combined data of the Daya Bay and RENO experiments. The 1-, 2-, and 3-$\sigma$ regions are shown with orange, blue, and purple, respectively. In the left panels only $\re[S]$ (or $\im[S])$ is varied at a time, while in the right panels both vary simultaneously. 
}}
\end{figure}
We are ready to extract constraints on the mixing angle $\theta_{13}$ and NSI parameters appearing in~\eref{OSC_generaltemplate} from a combination of  Daya Bay and RENO data.\footnote{For completeness, results using separate Daya Bay and RENO data are presented in \aref{separate}.}
In our analysis we do not treat $\Delta m^2_{31}$ as a free parameter, but rather use the best-fit value $\Delta m^2_{31}=2.52\times 10^{-3}$ eV$^2$~\cite{Esteban:2018azc} assuming normal ordering. 
This is justified because that result is dominated by other oscillation experiments than the reactor ones, and the new physics effect on the best-fit value and error is expected to be negligible.    
We have checked the stability of our results by letting $\Delta m^2_{31}$ vary within its $1\sigma$ uncertainty.

Consider first the case when only left-handed NSI are present: $\epsilon_L \neq 0$, $\epsilon_{R,S,P,T} = 0$. 
This corresponds to vanishing $\alpha$ and $\beta$ parameters in \eref{OSC_generaltemplate} and the only free parameter that remains is  $\tilde\theta_{13}$. 
As explained previously, the reactor experiments alone are not sensitive to new physics parametrized by $[\epsilon_L]_{e\mu}$ and $[\epsilon_L]_{e\tau}$, as these parameters can be absorbed into the unknown mixing angle $\theta_{13}$, and only the $\tilde \theta_{13}$ combination defined in  \eref{OSC_generaldictionary} is probed.  
After marginalizing the $\chi^2$ with respect to all the pull parameters we find the following result (all the uncertainties in this section are 68\%~CL):
\beq
\label{eq:FIT_sm} 
\sin^2 (2\tilde\theta_{13})=0.0841\pm0.0027. 
\eeq 
In this simple case, leaving $\Delta m^2_{31}$ as a free parameters would have a negligible impact on the confidence interval.

Next, we allow the scalar NSI to be non-zero: $\epsilon_S \neq 0$, $\epsilon_{R,P,T} = 0$.  
This implies $\alpha_P= \beta_P= 0$ in \eref{OSC_generaltemplate}, but now $\alpha_D$ and $\beta_D$ can be non-zero, that is to say, NSI effects can appear at the detection side.
Of course, now  we cannot use the value for $\tilde \theta_{13}$ in \eref{FIT_sm}, as it was obtained under assumption that $\epsilon_S = 0$. 
Instead,  we need to derive simultaneous constraints on $\tilde \theta_{13}$ and the combination of NSI parameters $[S]$ defined in \eref{OSC_ST}. 
We consider three different cases: 1) only $\re [S]$ is non-zero, 2)  only  $\im [S]$ is non-zero, and 3)  both are non-zero and independent.  
We present our results in \fref{SALL}. 
The orange, blue, and purple contours are the 1-, 2-, and 3-$\sigma$ allowed regions, respectively. 
For the real part we see some degeneracy between $\tilde\theta_{13}$ and $\re[S]$.  
This is expected: while the two lead to a different energy dependence of the survival probability in Eq.~(\ref{eq:OSC_generaltemplate}), they carry the same oscillatory dependence on  $L/E_\nu$.    
Setting $\im[S] = 0$ (upper left panel in \fref{SALL}), we find the following constraint
\beq
\re[S] = 0.54\pm0.39\,. 
\eeq 
The bounds are very loose, and the validity of the EFT expansion is not assured for the values of $\epsilon_S$ within these confidence intervals; in particular taking into account  ${\cal O}(\epsilon_S^2)$ terms in the survival probability may significantly change the results. 
The same holds for the validity of the expansion carried out to obtain the oscillation formula in \eref{OSC_generaltemplate}.
We conclude that $\re [S]$ cannot be reliably constrained by the existing oscillation data from reactor experiments.  
Moreover, in the presence of sizable $\re [S]$ the constraints on the effective mixing angle $\tilde \theta_{13}$ can be considerably relaxed. 

The situation is somewhat better for $\im [S]$ for which we can obtain $\cO(0.1)$ constraints. 
Assuming  $\re [S] = 0$ (upper right panel in \fref{SALL}) we find:  
\beq 
\im[S] =0.04\pm0.13 \,.
\eeq 
Comparing the upper and lower rows of Fig.~\ref{fig:SALL} we see that leaving $\re[S]$ as a free parameter to be marginalized over weakens the constraints on $\tilde \theta_{13}$, however for $\im[S]$ only the central value is slightly affected.  
In the situation where both  $\re[S]$ and $\im[S]$ are free parameters we find the constraints
\beq
\re[S]=0.95\pm0.37\, , \qquad   \im[S] = 0.08\pm0.14\, . 
\eeq 
Let us note that the 1-$\sigma$ region of $\re[S]$ is outside the validity range of the effective theory. We emphasize that the sign of the best fit value for $\im[S]$ depends on choosing the mass ordering, and would be flipped for the inverted ordering.

\begin{figure}[!tb]
\begin{center}
\includegraphics[width=0.97\textwidth]{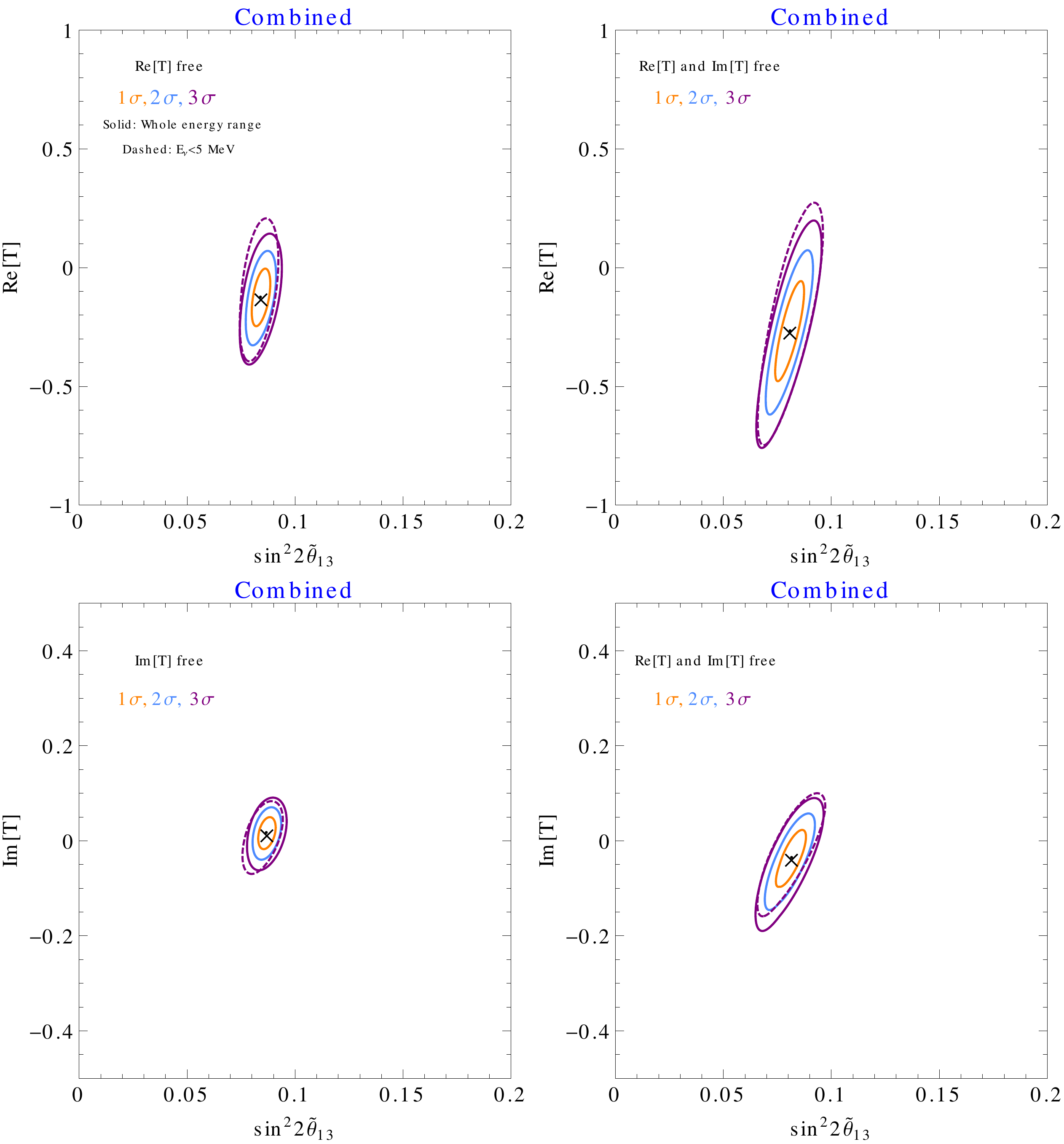}
\end{center}
\caption{
\footnotesize{
Allowed regions in the $(\sin^22\tilde\theta_{13}-\re[T])$ (first row) and $(\sin^22\tilde\theta_{13}-\im[T])$ plane (second row)  for the combined data of the Daya Bay and RENO experiments. The 1-, 2-, and 3-$\sigma$ regions are shown with orange, blue, and purple, respectively. The best fit values are marked by $\times$. In the left panels only $\re[T]$ $(\im[T])$ is varied at a time, while in the right panels both vary simultaneously.
The dashed curves correspond to the 3-$\sigma$ regions in the analysis where only events with $E_\nu<5$~MeV are taken into account. We note that the y-axis range is different in the upper and lower panels.}
\label{fig:TALL}
}
\end{figure}
Finally, we allow tensor NSI to be non-zero: $\hat \epsilon_T \neq 0$, $\epsilon_{R,S,P} = 0$.  
The effects of tensor NSI appear on both the production and detection sides. 
We consider again three distinct cases, one where only $\re [T]$ is non-zero, another where only  $\im [T]$ is non-zero, and  where both are free parameters.  
Before presenting our results we  recall that for calculating the production coefficients in Eq.~(\ref{eq:prodCoef}) we have assumed that the beta transitions in the reactors are of the Gamow-Teller type, while in fact almost $30\%$ of the decays are first-forbidden transitions~\cite{Hayes:2016qnu}. 
These transitions are expected to be more important at the high end of the neutrino spectrum~\cite{Hayen:2018uyg}. 
Therefore, to test the robustness of our conclusions, we compare the results obtained using the whole neutrino spectrum with the ones where the neutrino energies are restricted to $E_\nu<5$ MeV. 
We show the results in \fref{TALL}, using the same color coding as in \fref{SALL}. 
For the $3\sigma$ regions, we show the results using the entire neutrino energy spectrum  (solid contours) and with the $5$ MeV cut (dashed contours). 
The cut has limited impact on the preferred parameter regions, which suggests that the presence of forbidden transitions in the reactors should not affect our EFT constraints significantly.  
In this case the sensitivity to  $\re[T]$ is much better than for  $\re[S]$, with the $1\sigma$ contours contained within the validity regime of the EFT.
Using the entire neutrino spectrum  we find the following results for $\im[T]=0$:
\beq 
\re[T] = -0.124\pm0.081 \, , 
\eeq 
which changes to $-0.079 \pm 0.090$ after imposing the $E_\nu<5$ MeV cut.
For $\re[T] = 0$ the constraint on $\im [T]$ is 
\beq
\im[T] =-0.003\pm0.043 \, ,
\eeq 
or $\im[T] = -0.030\pm0.045$ after the $E_\nu<5$ MeV cut. 
One sees that the nonstandard term $\im[T]$ is the most strongly constrained one by reactor data.
When both $\re[T]$ and $\im[T]$ are free parameters we find the constraints \beq
\re[T] =-0.26\pm0.14 \, , \qquad 
\im[T] =-0.034\pm0.042  \,  , 
\eeq
relaxed to $-0.22\pm0.15$, $-0.084\pm0.042$
after the $E_\nu<5$ MeV cut is imposed.
Note that the errors for $\im[T]$  are a factor of three smaller than those for $\im [S]$.  

The last logical step would be to obtain a 5D likelihood function simultaneously for $\tilde \theta_{13}$, $\re [S]$, $\im [S]$, $\re [T]$, $\im [T]$ treated as independent free parameters. 
However, in this general case parameter degeneracies are probably too important to obtain meaningful constraints on the NSI parameters.
One could also try to derive constraints on the combination 
$[P] \equiv   e^{i \delta_{\rm CP}}  \left (s_{23} [\epsilon_P]_{e \mu} + c_{23} [\epsilon_P]_{e \tau}  \right )$ of the pseudo-scalar couplings in the effective Lagrangian. 
However, since the effects of $\epsilon_P$ are velocity-suppressed in our approximation, our analysis would be sensitive only to $[P] \gtrsim 1$, outside the validity range of the effective theory.     
As we discuss below in \sref{lpd}, much stronger constraints on $\epsilon_P$ can be derived from pion decays.

It is worth mentioning that our constraints on $\im [S]$ and $\im [T]$ are dominated by $\chi^2_{\rm spectral}$, with $\chi^2_{\rm rate}$ having a small impact on the confidence intervals. On the other hand, using only the spectral information we find that the degeneracy between $\re [S]$  and $\tilde\theta_{13}$ is worsened, which translates in weaker marginalized bounds. 
Last, we also note that the $\cO{(0.1-0.4)}$ bounds on $\re [S]$ and $\re [T]$ obtained above do not do justice to the sensitivity and potential of these measurements. One should keep in mind the large correlation with $\tilde{\theta}_{13}$ shown in Fig.~\ref{fig:SALL}. This translates to much more precise measurements in (i) less general scenarios, like the SM case in~\eref{FIT_sm} shows; or (ii) after combination with other measurements that are sensitive to the same coupling with different correlation.

\section{Non-oscillation constraints on EFT parameters}
\label{sec:non-osc}
\setcounter{equation}{0}

In the previous section we derived  a-few-percent-level constraints on
linear combinations of Wilson coefficients $[\epsilon_{S}]_{e\alpha}$ and $[\hat \epsilon_{T}]_{e\alpha}$, $\alpha = \mu, \tau$, from neutrino oscillations in reactor experiments. 
To see these results in a wider context, in this section we discuss precision observables that do not involve neutrino oscillations but are sensitive to the same parameters. 
There is one important difference between these two classes.
While the oscillations are sensitive to {\em linear} effects  $[\epsilon_{X}]_{e\alpha}$, the observables discussed below are sensitive to absolute values squared of these parameters (or their combinations).  
One consequence is that they cannot distinguish between real and imaginary parts. 
Furthermore, the dependence on  $[\epsilon_{X}]_{e\alpha}$ enters at ${O}(\Lambda^{-4})$ in the SMEFT expansion. 
It is in principle possible that their effects cancel against linear effects in $[\epsilon_{X}]_{ee}$, coming from dimension-6 or dimension-8 SMEFT operators.
For illustration we neglect such terms in the discussion below and set bounds when only one $[\epsilon_{X}]_{e\alpha}$ term is present in the Lagrangian at a time.
Because of these assumptions, the bounds obtained in this section are less robust than the ones from oscillations. 
Although they may be valid for the SMEFT derived from particular UV models,
it is important to keep in mind that they can be relaxed significantly if several interactions are present at the same time. 
A thorough analysis of such scenarios is however beyond the scope of this work, which is focused on neutrino physics.

\subsection{Neutron and nuclear beta decay}

Instead of the plethora of $\beta$-decay transitions happening inside nuclear reactors, one can search for nonstandard effects in specific decays that happen to be very clean both experimentally and theoretically~\cite{Gonzalez-Alonso:2018omy}. 
One expects strong bounds on nonstandard interactions involving wrong-flavor neutrinos from such studies, which has been used sometimes in the past as an argument to neglect e.g. scalar and tensor interactions.
However, to best of our knowledge, all available beta-decay analysis have focused on interactions involving electron neutrinos. 
 We amend this situation here, deriving the bounds on scalar and tensor operators with a wrong-flavor neutrino. Our experimental input are the so-called ${\cal F}t$ values of $0^+\to 0^+$ transitions~\cite{Hardy:2014qxa} and neutron data (lifetime and correlation coefficients). We use the same statistical approach and dataset as in the recent review in Ref.~\cite{Gonzalez-Alonso:2018omy} (Table 4, 5 and 7), but also including the recent PERKEO-III measurement of the beta asymmetry in neutron decay, $A_n=−0.11985(21)$~\cite{Markisch:2018ndu}. We note that the errors of the average lifetime and beta asymmetries are re-scaled {\it \`a la} PDG to take into account tensions among various measurements. 
Assuming only one interaction is present at a time we find the following 90\% CL bounds:
\beq
|[\epsilon_S]_{e\alpha}| \le 6.4\times 10^{-2}~,\qquad 
|[\hat \epsilon_T]_{e\alpha}| \le 4.4\times 10^{-2} \, . 
\eeq 
We note that the observables depend quadratically on these WEFT coefficients and thus the error distribution is highly non-gaussian.

\subsection{CKM unitarity}
In the beta decay fit discussed above one extracts {\it simultaneously} the non-standard scalar or tensor coupling and the SM parameters ($|V_{ud}|$ and the axial charge $g_A$). 
It is important to note that significant correlations between the scalar coupling and $|V_{ud}|$ appear. Thus, adding to this analysis the very precise $|V_{ud}|$ value obtained from CKM unitarity: $|V_{ud}^{unit.}|\equiv \left(1 - |V_{us}|^2 - |V_{ub}|^2 \right)^{1/2}$, has a drastic impact on the bound on scalar interactions~\cite{Gonzalez-Alonso:2018omy}. Namely:
\beq
\label{eq:epsilonSckm}
|[\epsilon_S]_{e \alpha}| \le 2.0\times 10^{-2}~\rm{(90\%\,CL)}~,
\eeq
where we used $V_{us}=0.2243(5)$
and $V_{ub}=0.00394(36)$~\cite{Tanabashi:2018oca}. 
Roughly speaking, the bound above comes from the comparison of $|V_{ud}|^2 \left(1 + g_S^2 |[\epsilon_S]_{e\alpha}|^2 \right)$, extracted from $0^+\to 0^+$ transitions, and $|V_{ud}^{unit.}|^2$. The former happens to be currently a bit smaller than the latter, which translates in a bound on $|[\epsilon_S]_{e\alpha}|$ more stringent than naively expected (because this interaction can only contribute positively).\footnote{In order to extract $|V_{ud}|$ from $\beta$ decays we have used the traditional values for the radiative corrections~\cite{Hardy:2014qxa,Marciano:2005ec}. Recently, Refs.~\cite{Seng:2018yzq,Seng:2018qru} presented new values for various corrections. These values give an even smaller value for $|V_{ud}|^2$, which translates into stronger constraints for the operators discussed in this work.}

CKM unitarity constrains also the offdiagonal vector coefficients, $[\epsilon_{L,R}]_{e\alpha}$. 
In a one-operator analysis we find 
\bea
\label{eq:epsilonLRckm}
|[\epsilon_{L,R}]_{e\alpha}| \le 1.9\times 10^{-2}~\rm{(90\%\,CL)}~.
\eea
The stringent bounds in~\eref{epsilonSckm} and~\eref{epsilonLRckm} assume  (i) the absence of other nonstandard $\beta$-decay couplings, as in the previous section; 
(ii) the absence of new physics effects in the extraction of $V_{us}$, $V_{ub}$, and the Fermi constant $G_F$;
and 
(iii) the 3-family setup, which is not an extra assumption in the SMEFT. 

\subsection{Leptonic pion decays}
\label{sec:lpd} 

The $\pi\to e\nu_e$ channel is extremely sensitive to pseudo-scalar couplings because the latter do not suffer the strong chiral suppression of the SM contribution (the SM width vanishes for zero electron mass). For a pseudo-scalar interaction with a wrong-flavor neutrino, a bound on $|[\epsilon_P]_{e\alpha}(\mu=2\,\rm{GeV})|$ can be derived from the clean ratio $R_\pi$ of the $\pi\to e\nu_e$ and $\pi\to \mu\nu_\mu$ widths~\cite{Britton:1992pg,Czapek:1993kc,Cirigliano:2007xi}.
Using the experimental and SM values $R_\pi =  1.2327(23) \times
10^{-4}$~\cite{Tanabashi:2018oca} and $R_\pi^{\rm SM} =  1.2352(1) \times 10^{-4}$~\cite{Cirigliano:2007xi}, the 90\%~CL constraint is given by 
\bea
\label{eq:epsilonPleptonicpiondecay}
|[\epsilon_P]_{e\alpha}|_{\mu=2\,\rm{GeV}} 
&\le&  7.5 \times 10^{-6}. 
\eea

BSM models generating scalar/tensor interactions often generate pseudoscalar interactions of similar magnitude. More importantly, even if this is not the case at tree-level, the pseudoscalar interactions is generated radiatively~\cite{Voloshin:1992sn}. For instance, the connection between $\epsilon_P (2\,\rm{GeV})$ and the coefficients at the EW scale and 1 TeV are given by~\cite{Gonzalez-Alonso:2017iyc}\footnote{In this expression $\epsilon_{S,P,T}(1\,\rm{TeV})$ should not be understood as a WEFT coefficient (the WEFT is not valid above $\mu \simeq m_Z$), but as a short notation for a linear combination of SMEFT coefficients, as given in \eref{EFT_wefttosmeft}.}
\bea
\epsilon_P (2\,{\rm GeV}) 
&=& 2.5\times 10^{-6}\,\epsilon_S (M_Z) + 1.7\,\epsilon_P (M_Z) -  0.0061 \,\hat \epsilon_T (M_Z) ~\\
&=& 0.0086\,\epsilon_S (1\,{\rm TeV}) + 2.1\,\epsilon_P (1\,{\rm TeV}) 
- 0.087\,\hat \epsilon_T (1\, {\rm TeV})~.
\eea
The larger mixing found in the 1-TeV case is due not only to the trivial larger running but also because the mixing happen to be larger in the SMEFT than in the WEFT~\cite{Gonzalez-Alonso:2017iyc}.

In full generality $\epsilon_P (2\,\rm{GeV})$ represents a different direction in the parameter space with respect to $\epsilon_{S,T} (2\,\rm{GeV})$. 
However, the mixing relations given above imply strong constraints on the tensor coupling in simplified scenarios where (pseudo)scalar and tensor couplings are not independent degrees of freedom, since severe cancellations among them are not possible anymore. 
For example, if we assume that the pseudo-scalar operator is not generated at tree-level at the high-scale, we obtain the following 90\% CL bounds
\bea
\left |[\hat \epsilon_T]_{e\alpha} + 3\times 10^{-4} [\epsilon_S]_{e\alpha}\right |_{\mu=2\,\rm{GeV}} &\leq & 1.0 \times 10^{-3}~\text{(running from $\mu=M_Z$)}~,
\nnl 
\left |[\hat \epsilon_T]_{e\alpha} - 4 \times 10^{-2} [\epsilon_S]_{e\alpha}\right |_{\mu=2\,\rm{GeV}} &\leq & 7.0 \times 10^{-5}~\text{(running from $\mu=1$~TeV)}~,
\eea
where the ranges correspond to the one-operator and global analysis discussed in~\eref{epsilonPleptonicpiondecay}. Let us note that the derivation of these bounds only takes into account log-enhanced one-loop corrections and it can be altered by finite pieces, especially if the running is not carried out to very high-energy scales.

\subsection{LHC ($pp\to e+\rm{MET}+X$)}
One can look for the same nonstandard charged-current interactions (or more precisely, for their SMEFT counterparts) in the Drell-Yan process $pp\to e+\rm{MET}+X$~\cite{Bhattacharya:2011qm,Cirigliano:2012ab}.
This connection requires additional assumptions such as the validity of the SMEFT at such high-energies, approximating $V_{ij}=\delta_{ij}$ in the SMEFT-WEFT mapping, and, especially, neglecting the contributions from dim-8 operators (since LHC bounds are dominated or very sensitive to dim-6 squared contributions).

Chirality-flipping interactions do not interfere with the SM and then the usual bounds on operators involving electron neutrinos actually to the incoherent sum over all three flavor neutrinos. 
Thus we can reinterpret the results from Fig.~8 of Ref.~\cite{Gupta:2018qil}, which used the 13-TeV ATLAS search with 36 fb$^{-1}$~\cite{Aaboud:2017efa}:
\bea
\left(  \sum_\alpha |[\epsilon_S]_{e\alpha}|^2 \right)^{1/2}  
\lesssim 2\times 10^{-3}~, \qquad
\left( \sum_\alpha |[\hat \epsilon_T]_{e\alpha}|^2 \right)^{1/2}  
\lesssim 2\times 10^{-3}~,
\eea
at 90\% CL and at $\mu=2$ GeV.

\subsection{Charged-lepton-flavor violation}

In the SMEFT, dimension-six operators that give rise to charged-current interactions between quarks and leptons also yield neutral-current interactions between quarks and pairs of charged leptons $\ell = e,\mu,\tau$. 
In consequence, neutrino interactions parametrized by off-diagonal $[\epsilon_X]_{e \alpha}$ appear in the Lagrangian together with 4-fermion  charged-lepton-flavor violating (CLFV) interactions. 
The latter mediate at tree- or loop-level such processes as $\ell \to \ell' \gamma$, $\ell \to 3 \ell'$, or $\ell N \to \ell' N$,  which for $\ell \neq \ell'$ have not been observed so far and are stringently constrained by experiment.  
The contribution of dimension-6 operators to CLFV observables arises at  $\cO(\Lambda^{-4})$, which is the leading order in this case because the SM contributions are absent.
The resulting constraints on lepton-flavor off-diagonal SMEFT operators are typically very severe \cite{Pruna:2015jhf,Crivellin:2017rmk,Davidson:2018rqt,Coy:2018bxr}. 
Using the analytical formula for the $\mu \to e$ conversion rate in Ref.~\cite{Cirigliano:2009bz} and the experimental bound 
${\rm Br}(\mu \to e)_{\rm Au} \leq 7 \times 10^{-13}$~\cite{Bertl:2006up}, one can constrain the SMEFT operators $[O^{(1)}_{lequ}]_{\mu e 11}$ and  $[O_{ledq}]_{\mu e 11}$. If their sum is the only nonzero term one finds the 90\%~CL bound
\beq 
\left |  [\epsilon_{S}]_{e \mu} \right | \lesssim 3 \times 10^{-6} \, .
\eeq 
Furthermore, non-observation of $\tau \to e \pi^+ \pi^-$ sets stringent constraints~\cite{Celis:2014asa} on scalar CLFV interactions involving  an electron and a tau. Assuming once again that only the low-energy scalar coupling $[\epsilon_{S}]_{e \tau}$ is generated, we find  the 90\%~CL bound
\beq 
\left |[\epsilon_{S}]_{e \tau} \right | \lesssim 4 \times 10^{-4} \, .
\eeq 
When such stringent limits on $[\eps_X]_{e \alpha}$ hold, observation of the scalar NSI effects in neutrino oscillation experiments is of course impossible. 
On the other hand, we are not aware of similar tree-level bounds in the literature on the SMEFT operator  $[O^{(3)}_{lequ}]_{\alpha e 11}$ contributing to the tensor  parameters $[\epsilon_{T}]_{e \alpha}$.  
We note that the CLFV constraints would not hold if the WEFT were {\em not } UV-completed by the SMEFT (because new physics is not much heavier than $m_Z$, or because electroweak symmetry is realized non-linealy in the UV theory), as then off-diagonal $\epsilon_X$ are not correlated in general with CLFV interactions.

\section{\label{sec:conclusion}Conclusions}

We have proposed a systematic approach to neutrino oscillations in the situation when neutrino interactions with matter are modified by heavy new physics. 
To this end we employed the model-independent framework of the SMEFT, with the Lagrangian organized into an expansion in powers of $1/\Lambda$, where $\Lambda$ is the mass scale of new particles affecting the neutrino interactions. 
The SMEFT framework enables consistent power-counting of new physics effects and identifying the leading order corrections to the neutrino oscillations probability.
In this paper we applied it to oscillations in short-baseline reactor experiments, however the formalism can be readily extended to other types of neutrino experiments.  

We calculated the survival probability of reactor antineutrinos at $\cO(\Lambda^{-2})$ in the SMEFT expansion, that is including linear effects of dimension-6 operators.  
The main result of this paper is given in Eq.~(\ref{eq:OSC_generaltemplate}), from which the dependence of the survival probability on the PMNS parameters and dimension-6 Wilson coefficients can be read off.  
We have taken into account all SMEFT operators that contribute at the leading order. 
In addition to operators affecting the SM-like (V-A) charged-current interactions between quarks and leptons, those mediating  scalar and tensor contact interactions contribute at the same order $\Lambda^{-2}$.
The latter lead to a different energy dependence of the neutrino production and detection amplitudes, which is reflected in Eq.~(\ref{eq:OSC_generaltemplate}).  
We also paid due attention to the interplay between the effects of dimension-6 operators and of the PMNS mixing. 
It is a familiar fact in electroweak precision measurements and flavor physics that some dimension-6 corrections to physical observables can be absorbed into SM parameters, such as the Higgs vacuum expectation value or the Wolfenstein parameters.  
In the present case one observes an analogous effect. 
It turns out that corrections to the survival probability due to lepton-flavor off-diagonal V-A interactions parametrized by $[\epsilon_L]_{e \alpha}$ 
can be absorbed into a redefinition of the mixing angle $\theta_{13}$~\cite{Ohlsson:2008gx}. 
Since $\theta_{13}$ is not known a-priori (other than from the very reactor experiments we consider here), the existing data only constrain a linear combination $\tilde \theta_{13}$ of the original mixing angle $\theta_{13}$ and dimension-6 Wilson coefficients, cf. \eref{OSC_generaldictionary}, but not the two separately. 
We conclude that,  at the leading order in the SMEFT expansion, reactor neutrino experiments alone only constrain the scalar and tensor interactions, but not the V-A ones.
This explains the origin of the degeneracy between $\theta_{13}$ and off-diagonal V-A NSI found in the previous literature.

We pointed out that neutrino oscillations can probe, at the linear (order $\Lambda^{-2}$) level, the dimension-6 tensor and scalar SMEFT operators that are off-diagonal in the lepton-flavor space.
To our understanding, it is the unique class of observables where this is the case. 
Consequently, the oscillation constraints on these operators are robust as long as the expansion of the SMEFT Lagrangian in powers of $1/\Lambda$ is quickly convergent.
The same operators can be probed by meson and nuclear decays or by production processes at the LHC, however in those cases they enter quadratically (at order $\Lambda^{-4}$). 
Such constraints are then subject to model-dependent assumptions about other dimension-6 and dimension-8 contributions to the same observables, and are thus less robust.  

We identified the linear combinations of Wilson coefficients of scalar and tensor SMEFT operators that can be constrained by reactor oscillation experiments, cf. \eref{OSC_ST}. 
Using the most recent data from Daya Bay and RENO, we derived numerical constraints on these combinations.  
Under various more or less constraining assumptions, they are presented in \sref{fit} and illustrated in Figs.~\ref{fig:SALL} and~\ref{fig:TALL}. 
As of today, constraints at a few percent level can be extracted from the publicly available reactor experiment data. 
This is competitive with the constraints extracted from nuclear decays (which are less robust, as discussed above). 
At face value, the LHC constraints on the same operators are at least an order of magnitude more stringent. However, they are more model-dependent, and rely on the assumption that the SMEFT is a valid framework at TeV energy scales. 
We also  derived stringent constraints on scalar and, especially, tensor Wilson coefficients arising due to renormalization group mixing with the pseudoscalar ones. 
The latter are strongly constrained by pion decays thanks to chiral enhancement. 
Again, those constraints are less robust, in particular they depend on the starting point of the running,  and assume the absence of cancellations between different contributions.
Finally, CLFV processes typically place severe constraints on SMEFT operators that are off-diagonal in lepton-flavor indices.  
In particular, $\mu \to e$ conversion on nuclei strongly constrains operators contributing to the NSI parameter $[\epsilon_S]_{e \mu}$, 
while $\tau \to e \pi^+ \pi^-$ decays constrain the operators contributing  to $[\epsilon_S]_{e \tau}$. 
To our knowledge, however, CLFV constraints on the operators contributing  at tree level to the tensor parameters  $[\epsilon_T]_{e \mu}$,  and  $[\epsilon_T]_{e \tau}$ (that also affect reactor neutrino oscillations at the leading order) are not given in the literature. 

We note that the main goal of reactor experiments so far has been a precise determination of the mixing angle $\theta_{13}$ in the standard context, 
and the analyses were certainly not optimized for constraining SMEFT operators. 
We believe that with more data, more detailed spectral information, and targeted analyses, the constraints obtained in this paper can be significantly improved.
Furthermore, in our analysis the constraining power of reactor experiments is weakened by a partial degeneracy between NSI and the effective mixing angle $\tilde \theta_{13}$.  
Designing new observables sensitive to a {\em different} linear combination of  $\tilde \theta_{13}$ and $[\epsilon_X]_{e\alpha}$ may be another path to increasing senisitivity to new physics.

\section*{Acknowledgements}
ZT is supported by Funda\c{c}\~{a}o de Amparo \`{a} Pesquisa do Estado de S\~{a}o Paulo (FAPESP) under contract 16/02636-8. A.F. is partially supported by the European Union’s Horizon 2020 research and innovation programme under the Marie Sk\l{}odowska-Curie grant agreements No 690575 and No 674896. M.G-A. is supported by a Marie Sk\l{}odowska-Curie Individual Fellowship of the European Commission’s Horizon 2020 Programme under contract  number  745954  Tau-SYNERGIES.  
ZT would like to thank Joachim Kopp, Yuber F. Perez and Ivan Esteban for their valuable discussions and the hospitality of LPT Orsay and CERN Theory Department where this work was initiated and developed.

\appendix
\renewcommand{\theequation}{\Alph{section}.\arabic{equation}} 
\setcounter{equation}{0}

\section{WEFT without SMEFT}
\label{app:without}
\setcounter{equation}{0}

The WEFT is an effective theory below the scale $\mu \lesssim m_W$ describing interactions of  the SM particles with the exception of the $W$, $Z$  and Higgs bosons and the top quark, which have been integrated out.
In the main body of this paper we treated the WEFT as the low-energy theory of the SMEFT. 
However, the WEFT is a consistent EFT in its own right, which can be valid even if it is not completed by the SMEFT at higher energies. 
This caveat is relevant if the masses of BSM particles are between a few and a 100~GeV, or if the electroweak symmetry in the BSM theory is realized non-linearly.  
In this situation the leading order Lagrangian relevant for neutrino oscillations is still the one of \eref{EFT_lweft}, 
however the parameters $\epsilon_X$ are no longer related by matching to the SMEFT parameters at higher energies. 
The most important practical consequence is that $\epsilon_R$ may contain non-diagonal elements at the leading order in the WEFT. 
Then the reactor antineutrino survival probability in \eref{OSC_generaltemplate} is generalized to 
\bea
\label{eq:OSC_generaltemplate2}
P_{\bar \nu_e \to \bar \nu_e} (L,E_\nu)
  & =  & 
1 
- \sin^2 \left (\Delta m_{31}^2 L \over 4 E_\nu \right ) \sin^2 \left ( 2 \hat \theta_{13} -  \alpha_D  {m_e \over E_\nu - \Delta}  -   \alpha_P {m_e \over f_T(E_\nu)}  \right) 
\nnl 
&+ & \sin \left (\Delta m_{31}^2 L \over 2 E_\nu \right )   \sin (2 \hat \theta_{13})  \left ( \gamma_R + 
\beta_D {m_e \over E_\nu - \Delta} - \beta_P {m_e \over f_T(E_\nu)}  \right ) + \cO(\epsilon_X^2) , 
\eea 
where the definitions for $\gamma_R$ and the (new) effective mixing angle are
\bea
\hat \theta_{13}  &=&  \theta_{13} +  \re [L]  - {3 g_A^2 \over 3 g_A^2 +1} \re[R]\, ,\\
\gamma_R &=& - {2 \over 3 g_A^2 + 1} \im [R]~.
\eea 
The definitions of the coefficients $\alpha_{P,D}$, $\beta_{P,D}$ in~\eref{OSC_generaltemplate2} are given in \eref{OSC_generaldictionary}, whereas $[R]$ is defined in analogy to $[L]$, $[S]$, and $[T]$, i.e.
$[R] \equiv  e^{i \delta_{\rm CP}} \left (s_{23} [\epsilon_R]_{e \mu} + c_{23} [ \epsilon_R]_{e \tau}\right )$.

\begin{figure}[t]
\centering
\includegraphics[width=0.8\textwidth]{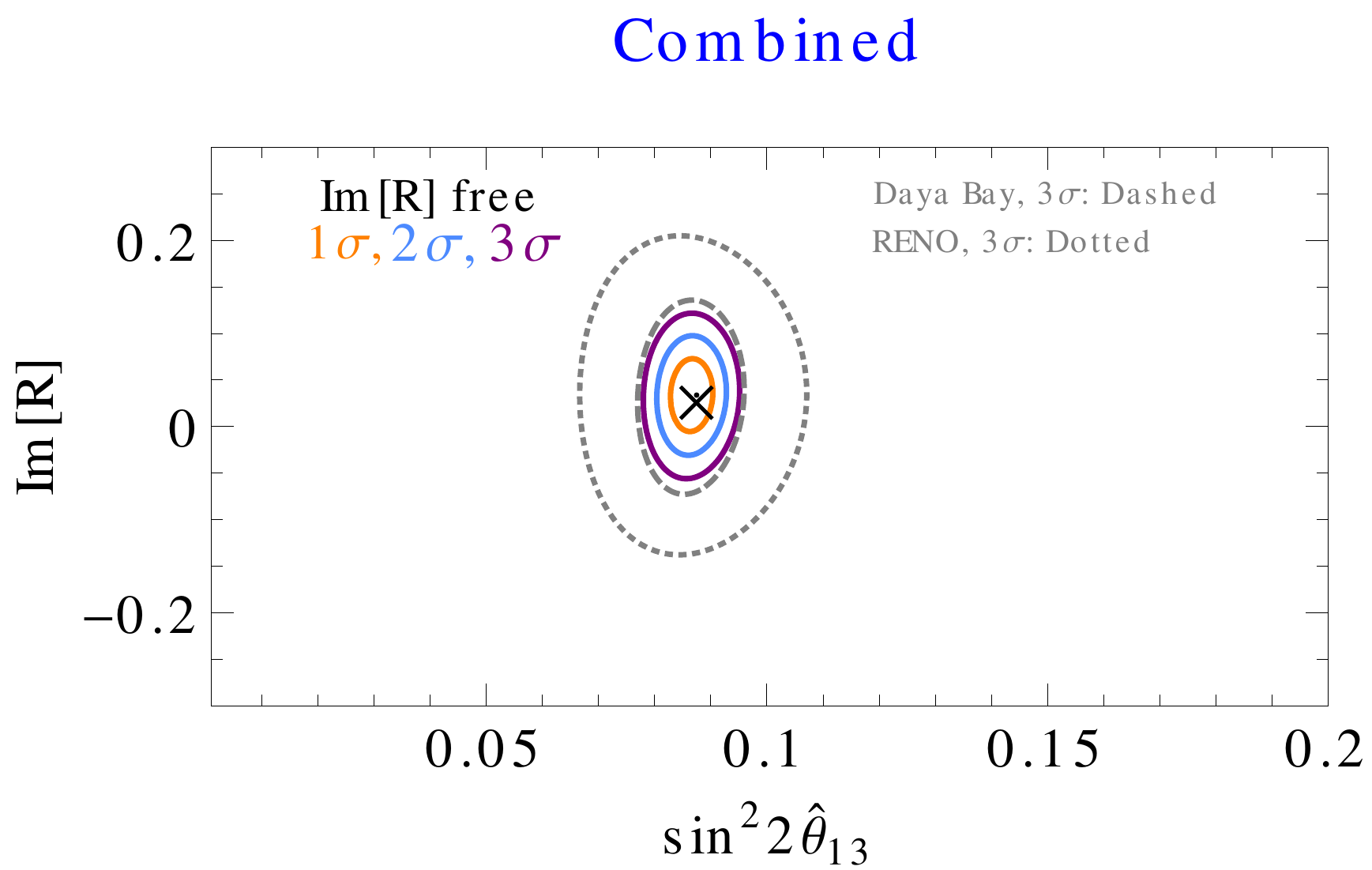}
\caption{\label{fig:ImR}
Allowed regions in the $(\sin^2 2\hat\theta_{13}-\im[R])$ plane for the combined data of the Daya Bay and RENO experiments. The 1-, 2-, and 3-$\sigma$ regions are shown with orange, blue, and purple, respectively. The Daya Bay and RENO $3\sigma$ regions are shown with the grey dashed and dotted curves, respectively. }
\end{figure}
In this setting the survival probability acquires dependence on one linear combination of the $\epsilon_R$ parameters, which is in principle distinguishable from other parameters in  \eref{OSC_generaltemplate2} due to the distinct $L/E_\nu$ and $E_\nu$ dependence.
We show in Fig.~\ref{fig:ImR} the results of the combined analysis of the Daya Bay and RENO experiments allowing only the right-handed NSI to be non-zero: $\epsilon_R\neq0,~\epsilon_{S,P,T}=0$. 
Under this assumption we find  
\beq 
\im[R] =0.034\pm0.026\,.
\eeq 

\section{Traditional NSI formalism}
\label{app:NSInu}
\setcounter{equation}{0}

In the neutrino literature,  NSI are typically parametrized by  the effective couplings $\epsilon^s$ and $\epsilon^d$, 
which correspond to non-standard effects in neutrino production and detection, respectively (see e.g. Ref.~\cite{Kopp:2007ne}). 
In this approach,  neutrinos produced at the source and detected at the detectors are not pure flavor states: 
 \bea
 \ket{\nu_\alpha^s}=\ket{\nu_\alpha}+\sum_{\beta=e,\mu,\tau} \epsilon^s_{\alpha\beta}\ket{\nu_\beta}\,,~~~~~~~~~~
 \bra{\nu_\beta^d}=\bra{\nu_\beta}+\sum_{\alpha=e,\mu,\tau} \epsilon^d_{\beta\alpha}\bra{\nu_\alpha}\,, 
 \eea
 where the first and second indices in $ \epsilon^s_{\alpha\beta}$ correspond to the flavors of the charged lepton and neutrino, respectively, which is reversed for $ \epsilon^d_{\alpha\beta}$. 
 The oscillation probability is given by 
 \bea
P_{\nu_\alpha^s\to\nu_\beta^d}=|\bra{\nu_\beta^d}e^{-iHL}\ket{\nu^s_\alpha}|^2\,,
 \eea
 where in the absence of NSI in propagation the Hamiltonian is given by
 \bea
 H_{\alpha\beta}=\frac{1}{2E_\nu}U_{\alpha j}
 \begin{pmatrix}
 0&0&0\\
 0&\Delta m^2_{21}&0\\
 0&0&\Delta m^2_{31}
 \end{pmatrix}
 U^{\dagger}_{k \beta}\,.
 \eea
 Up to first order in $ \epsilon$'s, 
  rewriting $ \epsilon^s_{e\alpha}=| \epsilon^s_{e\alpha}|e^{i\phi^s_{e\alpha}}$ and $ \epsilon^d_{\beta e}=| \epsilon^d_{\beta e}|e^{i\phi^d_{\beta e}}$, 
 and taking the limits $\Delta m_{21}^2 L/E_\nu \ll 1$, $\cos \theta_{13} \approx 1$, the survival probability of electron antineutrinos becomes \cite{Kopp:2007ne,Agarwalla:2014bsa}
  \bea\label{eq:NSIprob}
P_{\bar\nu^s_e\to\bar\nu^d_e}
=1&-&\sin^22\theta_{13}\sin^2\frac{\Delta m^2_{31}L}{4E_\nu}+2|  \epsilon_{ee}^s|\cos\phi^s_{ee}+2|  \epsilon_{ee}^d|\cos\phi^d_{ee}
\nonumber\\ &-&
4 s_{13}\sin^2\frac{\Delta m^2_{31}L}{4E_\nu}
\big [
s_{23}| {\epsilon}^s_{e \mu }|\cos(\delta_{\rm CP}-\phi^s_{e\mu }) 
+ s_{23}| {\epsilon}^d_{\mu e}|\cos(\delta_{\rm CP}+\phi^d_{\mu e})
+ (\mu \to \tau, s \to c)
\big ]
\nonumber\\ &+ &
2s_{13}\sin\frac{\Delta m^2_{31}L}{2E_\nu}
\big[
s_{23}| {\epsilon}^s_{e \mu }|\sin(\delta_{\rm CP}-\phi^s_{e\mu }) 
-s_{23}| {\epsilon}^d_{\mu e}|\sin(\delta_{\rm CP}+\phi^d_{\mu e})
+ (\mu \to  \tau, s \to c)
\big] \,,
\nonumber\\
 \eea
Comparing the probability in Eq.~(\ref{eq:NSIprob}) with the WEFT formula in Eq.~(\ref{eq:OSC_generaltemplate2}), the two agree given the matching between the effective couplings
\bea
(  \epsilon_{e \alpha}^s)^* &= & 
[\epsilon_L]_{e \alpha} - [\epsilon_R]_{e \alpha}
- {m_e \over f_T(E_\nu)} {g_T \over g_A}  [\hat \epsilon_T]_{e \alpha},
\nnl 
\epsilon_{\alpha e}^d  &= & 
[\epsilon_L]_{e \alpha} 
- {3 g_A^2 - 1 \over 3 g_A^2 + 1} [\epsilon_R]_{e \alpha} 
- {m_e \over E_\nu - \Delta}\Big( {g_S\over 3g_A^2 + 1}  [\epsilon_S]_{e \alpha} - {3 g_A g_T \over 3g_A^2 + 1}  [\hat \epsilon_T]_{e \alpha}\Big), 
\eea 
for $\alpha = \mu,\tau$. 
On the other hand, matching \eref{NSIprob} to Eq.~(\ref{eq:OSC_generaltemplate}) we find $\re \left(\epsilon_{ee}^s + \epsilon_{ee}^d \right )=0$.  
In our formalism, the diagonal EFT coefficients do not enter the survival probability, but rather the total rate of produced and detected neutrinos (see~\sref{observables}).
This way, $P_{\bar\nu^s_e\to\bar\nu^d_e}$ can indeed be interpreted as a probability, since non-oscillatory terms linearly  proportional to $\epsilon_{ee}^{s,d}$ would lead to  $P_{\bar\nu^s_e\to\bar\nu^d_e}>1$ for some choices of parameters. It is important to note that here we compare the ``traditional'' NSI formalism only to short baseline experiments, while the matter effects would still need to be included for long base-line experiments.

\section{Separate RENO and Daya Bay analyses}
\label{app:separate}
\setcounter{equation}{0}

For completeness, in this appendix we show separate RENO and Daya Bay constraints on the mixing angle and the NSI parameters in \eref{OSC_generaltemplate}. 
This allows us to compare the constraining power of the two experiments and their relative weight in the combined fit.  
We follow the same presentation as in \sref{fit}. 
We consider first the case when only left-handed NSI are present: $\epsilon_L \neq 0$, $\epsilon_{R,S,P,T} = 0$, in which case the only free parameter is  $\tilde\theta_{13}$. 
It is constrained as 
\beq 
\left(\sin^22\tilde\theta_{13}\right)_{\rm{Daya~Bay}} = 0.0841\pm 0.0028\, , \qquad 
\left(\sin^22\tilde\theta_{13}\right)_{\rm{RENO}} = 0.0843\pm 0.0064 \, . 
\eeq 
 These results are in good agreement with the best fit values end errors for  $\theta_{13}$ reported by the Daya Bay and RENO collaborations~\cite{Adey:2018zwh,Bak:2018ydk}.

\begin{figure}[!tb]
\begin{center}
\includegraphics[width=0.85\textwidth]{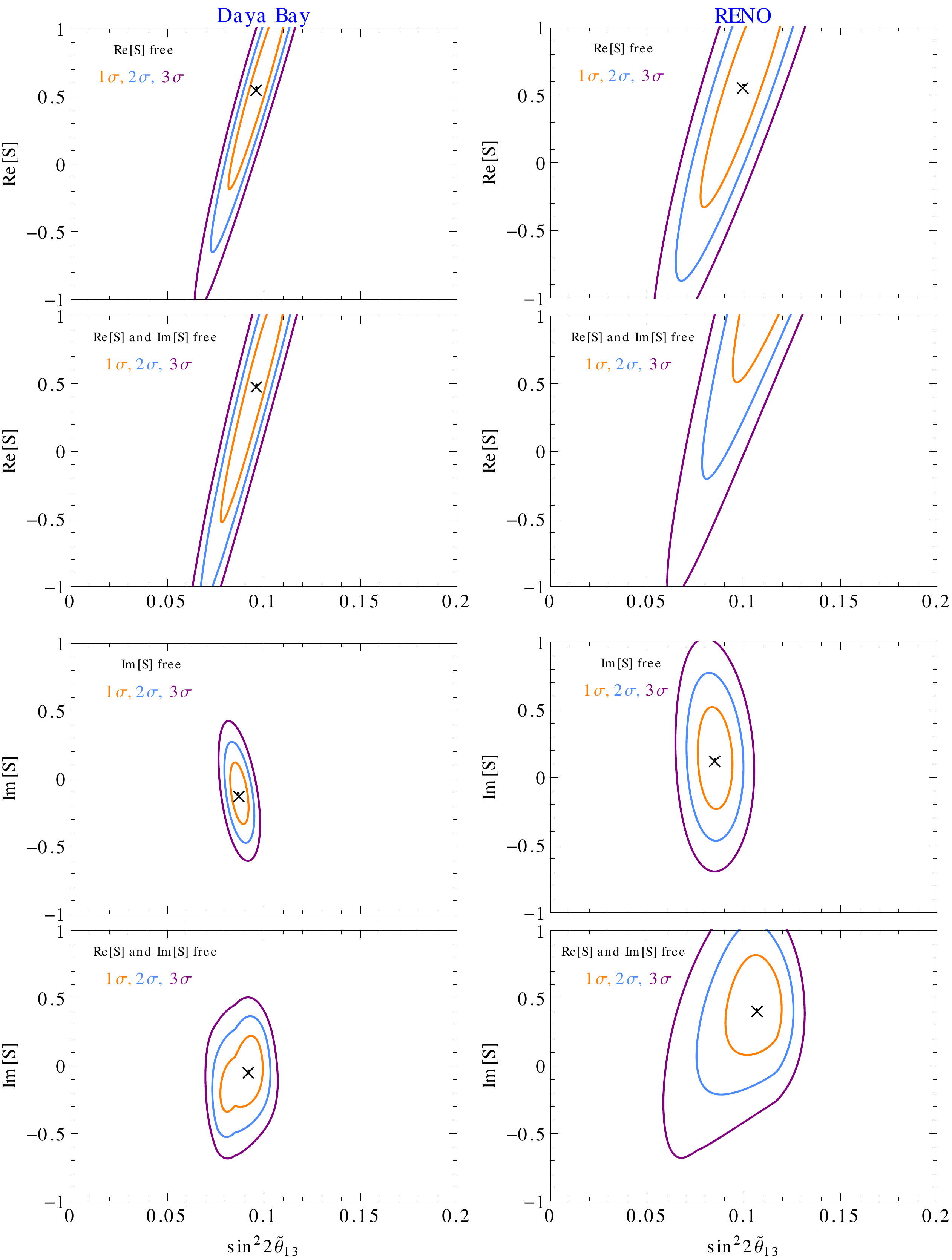}
\end{center}
\caption{
\footnotesize{Allowed regions in the $(\sin^22\tilde\theta_{13}-\re[S])$ (first and second rows) and $(\sin^22\tilde\theta_{13}-\im[S])$ plane (third and fourth rows), for the Daya Bay (left) and RENO (right) experiments. The 1-, 2-, and 3-$\sigma$ regions are shown with orange, blue, and purple, respectively. The best fit values are marked by $\times$.}
\label{fig:SALLcomparison}
}
\end{figure}

Next, we allow the scalar NSI to be non-zero: $\epsilon_S \neq 0$, $\epsilon_{R,P,T} = 0$.  
For $\im [S] = 0$ we find 
\beq 
\re[S]_{\rm{Daya~Bay}}=0.56\pm0.50 \,,
\qquad
 \re[S]_{\rm{RENO}}=-0.57\pm0.60 \, . 
\eeq 
For  $\re [S] = 0$ we find 
\beq 
\im[S]_{\rm{Daya~Bay}} = -0.11\pm0.15\,, \qquad 
\im[S]_{\rm{RENO}} =0.14\pm0.25\, . 
\eeq 
When both $\re [S]$ and $\im [S]$ are free parameters we find 
\begin{eqnarray} 
\re[S]_{\rm{Daya~Bay}}=0.50\pm0.69\,,  
 & \qquad &    
 0.87 <\re[S]_{\rm{RENO}} \,,
 \nnl 
\im[S]_{\rm{Daya~Bay}}=0.03\pm0.17\,,
 & \qquad &  
\im[S]_{\rm{RENO}} =0.45\pm{0.24}\,. 
\end{eqnarray}
\begin{figure}[!tb]
\begin{center}
\includegraphics[width=0.82\textwidth]{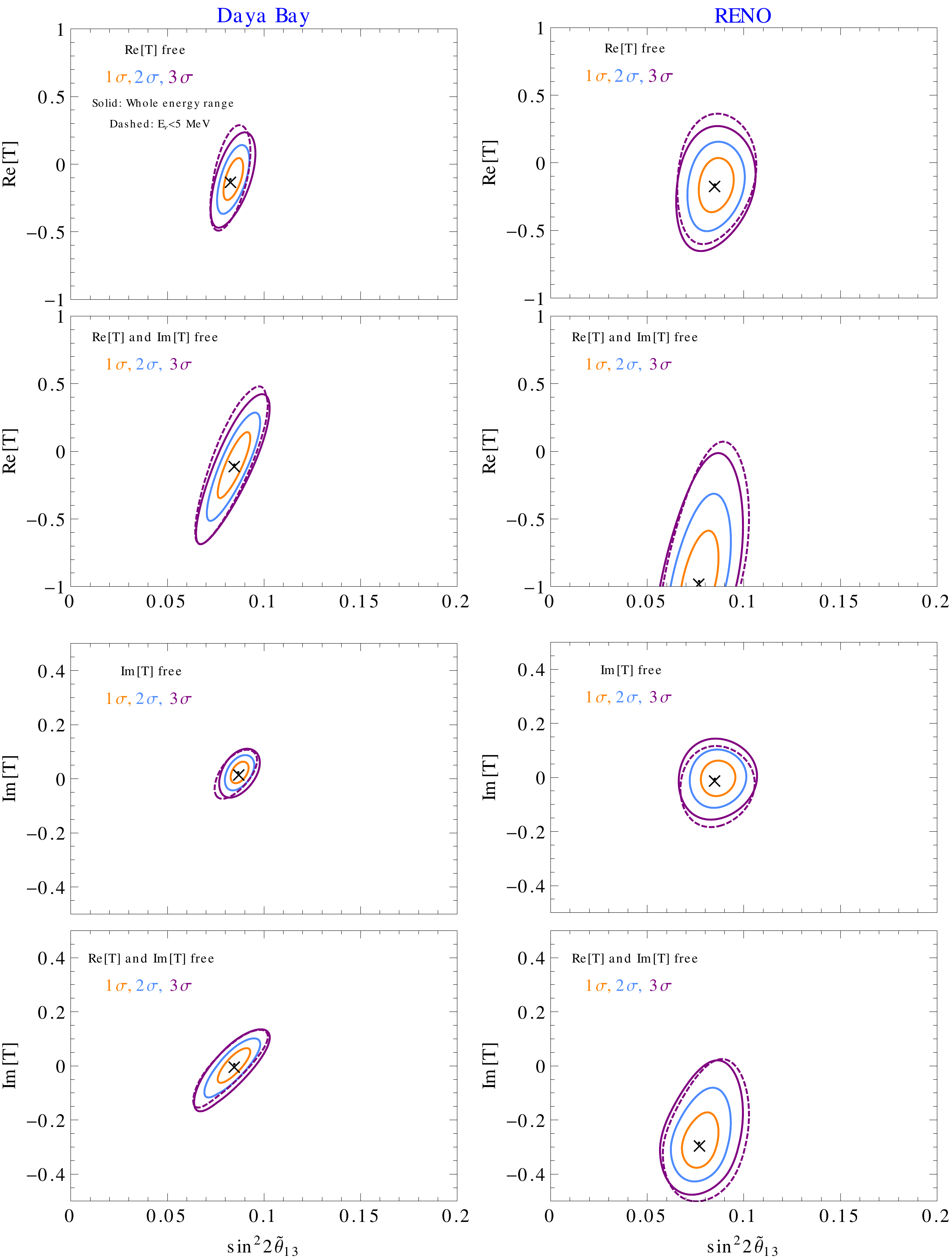}
\end{center}
\caption{
\footnotesize{
Allowed regions in the $(\sin^22\tilde\theta_{13}-\re[T])$ (first and second rows) and $(\sin^22\tilde\theta_{13}-\im[T])$ plane (third and fourth rows), for the Daya Bay (left) and RENO (right) experiments. The 1-, 2-, and 3-$\sigma$ regions are shown with orange, blue, and purple, respectively. The best fit values are marked by $\times$.
The dashed curves correspond to the analysis where only events with $E_\nu<5$~MeV are taken into account. We note that the y-axis range is different in the upper and lower panels.}
\label{fig:TALLcomparison}
}
\end{figure}

Finally, we allow tensor NSI to be non-zero: $\hat \epsilon_T \neq 0$, $\epsilon_{R,S,P} = 0$.  
For $\im [T] = 0$ we find 
\beq 
\re[T]_{\rm{Daya~Bay}} =-0.11\pm0.10 \, , \qquad 
\re[T]_{\rm{RENO}}=-0.08\pm0.14 \, . 
\eeq 
 For $\re [T] = 0$ we find 
 \beq
 \im[T]_{\rm{Daya~Bay}} = 0.023\pm0.026 \, , \qquad 
\im[T]_{\rm{RENO}}=- 0.003\pm0.043 \, . 
 \eeq 
 When both $\re [T]$ and $\im [T]$ are free parameters we find 
 \begin{eqnarray}
\re[T]_{\rm{Daya~Bay}}=-0.09\pm0.16 \, ,
& \qquad &  
\re[T]_{\rm{RENO}}=-0.91\pm0.25\, ,
 \nnl 
\im[T]_{\rm{Daya~Bay}}=0.005\pm0.041\, , 
& \qquad &  
\im[T]_{\rm{RENO}}=-0.028\pm0.069 \, . 
\end{eqnarray}

\bibliographystyle{./JHEP.bst} 
\bibliography{main}

\providecommand{\href}[2]{#2}\begingroup\raggedright\begin{thebibliography}{10}

\bibitem{Esteban:2018azc}
I.~Esteban, M.~C. Gonzalez-Garcia, A.~Hernandez-Cabezudo, M.~Maltoni, and
  T.~Schwetz, {\it {Global analysis of three-flavour neutrino oscillations:
  synergies and tensions in the determination of $\theta_{23}$, $\delta_{CP}$,
  and the mass ordering}},  {\em JHEP} {\bf 01} (2019) 106,
  [\href{http://arxiv.org/abs/1811.05487}{{\tt arXiv:1811.05487}}].

\bibitem{Antusch:2006vwa}
S.~Antusch, C.~Biggio, E.~Fernandez-Martinez, M.~B. Gavela, and J.~Lopez-Pavon,
  {\it {Unitarity of the Leptonic Mixing Matrix}},  {\em JHEP} {\bf 10} (2006)
  084, [\href{http://arxiv.org/abs/hep-ph/0607020}{{\tt hep-ph/0607020}}].

\bibitem{Kopp:2007ne}
J.~Kopp, M.~Lindner, T.~Ota, and J.~Sato, {\it {Non-standard neutrino
  interactions in reactor and superbeam experiments}},  {\em Phys. Rev.} {\bf
  D77} (2008) 013007, [\href{http://arxiv.org/abs/0708.0152}{{\tt
  arXiv:0708.0152}}].

\bibitem{Bolanos:2008km}
A.~Bolanos, O.~G. Miranda, A.~Palazzo, M.~A. Tortola, and J.~W.~F. Valle, {\it
  {Probing non-standard neutrino-electron interactions with solar and reactor
  neutrinos}},  {\em Phys. Rev.} {\bf D79} (2009) 113012,
  [\href{http://arxiv.org/abs/0812.4417}{{\tt arXiv:0812.4417}}].

\bibitem{Ohlsson:2008gx}
T.~Ohlsson and H.~Zhang, {\it {Non-Standard Interaction Effects at Reactor
  Neutrino Experiments}},  {\em Phys. Lett.} {\bf B671} (2009) 99--104,
  [\href{http://arxiv.org/abs/0809.4835}{{\tt arXiv:0809.4835}}].

\bibitem{Delepine:2009am}
D.~Delepine, V.~Gonzalez~Macias, S.~Khalil, and G.~Lopez~Castro, {\it {QFT
  results for neutrino oscillations and New Physics}},  {\em Phys. Rev.} {\bf
  D79} (2009) 093003, [\href{http://arxiv.org/abs/0901.1460}{{\tt
  arXiv:0901.1460}}].

\bibitem{Biggio:2009nt}
C.~Biggio, M.~Blennow, and E.~Fernandez-Martinez, {\it {General bounds on
  non-standard neutrino interactions}},  {\em JHEP} {\bf 08} (2009) 090,
  [\href{http://arxiv.org/abs/0907.0097}{{\tt arXiv:0907.0097}}].

\bibitem{Leitner:2011aa}
R.~Leitner, M.~Malinsky, B.~Roskovec, and H.~Zhang, {\it {Non-standard
  antineutrino interactions at Daya Bay}},  {\em JHEP} {\bf 12} (2011) 001,
  [\href{http://arxiv.org/abs/1105.5580}{{\tt arXiv:1105.5580}}].

\bibitem{Ohlsson:2012kf}
T.~Ohlsson, {\it {Status of non-standard neutrino interactions}},  {\em Rept.
  Prog. Phys.} {\bf 76} (2013) 044201,
  [\href{http://arxiv.org/abs/1209.2710}{{\tt arXiv:1209.2710}}].

\bibitem{Esmaili:2013fva}
A.~Esmaili and A.~{\relax Yu}. Smirnov, {\it {Probing Non-Standard Interaction
  of Neutrinos with IceCube and DeepCore}},  {\em JHEP} {\bf 06} (2013) 026,
  [\href{http://arxiv.org/abs/1304.1042}{{\tt arXiv:1304.1042}}].

\bibitem{Agarwalla:2014bsa}
S.~K. Agarwalla, P.~Bagchi, D.~V. Forero, and M.~Tortola, {\it {Probing
  Non-Standard Interactions at Daya Bay}},  {\em JHEP} {\bf 07} (2015) 060,
  [\href{http://arxiv.org/abs/1412.1064}{{\tt arXiv:1412.1064}}].

\bibitem{Ludl:2016ane}
P.~O. Ludl and W.~Rodejohann, {\it {Direct Neutrino Mass Experiments and Exotic
  Charged Current Interactions}},  {\em JHEP} {\bf 06} (2016) 040,
  [\href{http://arxiv.org/abs/1603.08690}{{\tt arXiv:1603.08690}}].

\bibitem{Choudhury:2018xsm}
D.~Choudhury, K.~Ghosh, and S.~Niyogi, {\it {Non-Standard Neutrino Interactions
  : Obviating Oscillation Experiments}},
  \href{http://arxiv.org/abs/1801.01513}{{\tt arXiv:1801.01513}}.

\bibitem{Heeck:2018nzc}
J.~Heeck, M.~Lindner, W.~Rodejohann, and S.~Vogl, {\it {Non-Standard Neutrino
  Interactions and Neutral Gauge Bosons}},  {\em SciPost Phys.} {\bf 6} (2019)
  038, [\href{http://arxiv.org/abs/1812.04067}{{\tt arXiv:1812.04067}}].

\bibitem{Altmannshofer:2018xyo}
W.~Altmannshofer, M.~Tammaro, and J.~Zupan, {\it {Non-standard neutrino
  interactions and low energy experiments}},
  \href{http://arxiv.org/abs/1812.02778}{{\tt arXiv:1812.02778}}.

\bibitem{AristizabalSierra:2018eqm}
D.~Aristizabal~Sierra, V.~De~Romeri, and N.~Rojas, {\it {COHERENT analysis of
  neutrino generalized interactions}},  {\em Phys. Rev.} {\bf D98} (2018)
  075018, [\href{http://arxiv.org/abs/1806.07424}{{\tt arXiv:1806.07424}}].

\bibitem{Coloma:2017ncl}
P.~Coloma, M.~C. Gonzalez-Garcia, M.~Maltoni, and T.~Schwetz, {\it {COHERENT
  Enlightenment of the Neutrino Dark Side}},  {\em Phys. Rev.} {\bf D96}
  (2017), no.~11 115007, [\href{http://arxiv.org/abs/1708.02899}{{\tt
  arXiv:1708.02899}}].

\bibitem{Esteban:2018ppq}
I.~Esteban, M.~C. Gonzalez-Garcia, M.~Maltoni, I.~Martinez-Soler, and
  J.~Salvado, {\it {Updated Constraints on Non-Standard Interactions from
  Global Analysis of Oscillation Data}},  {\em JHEP} {\bf 08} (2018) 180,
  [\href{http://arxiv.org/abs/1805.04530}{{\tt arXiv:1805.04530}}].

\bibitem{Bergmann:1999rz}
S.~Bergmann, Y.~Grossman, and E.~Nardi, {\it {Neutrino propagation in matter
  with general interactions}},  {\em Phys. Rev.} {\bf D60} (1999) 093008,
  [\href{http://arxiv.org/abs/hep-ph/9903517}{{\tt hep-ph/9903517}}].

\bibitem{Farzan:2017xzy}
Y.~Farzan and M.~Tortola, {\it {Neutrino oscillations and Non-Standard
  Interactions}},  {\em Front.in Phys.} {\bf 6} (2018) 10,
  [\href{http://arxiv.org/abs/1710.09360}{{\tt arXiv:1710.09360}}].

\bibitem{Adey:2018zwh}
{\bf Daya Bay} Collaboration, D.~Adey et~al., {\it {Measurement of electron
  antineutrino oscillation with 1958 days of operation at Daya Bay}},
  \href{http://arxiv.org/abs/1809.02261}{{\tt arXiv:1809.02261}}.

\bibitem{Bak:2018ydk}
{\bf RENO} Collaboration, G.~Bak et~al., {\it {Measurement of Reactor
  Antineutrino Oscillation Amplitude and Frequency at RENO}},
  \href{http://arxiv.org/abs/1806.00248}{{\tt arXiv:1806.00248}}.

\bibitem{Buchmuller:1985jz}
W.~Buchmuller and D.~Wyler, {\it {Effective Lagrangian Analysis of New
  Interactions and Flavor Conservation}},  {\em Nucl.Phys.} {\bf B268} (1986)
  621--653.

\bibitem{Grzadkowski:2010es}
B.~Grzadkowski, M.~Iskrzynski, M.~Misiak, and J.~Rosiek, {\it {Dimension-Six
  Terms in the Standard Model Lagrangian}},  {\em JHEP} {\bf 1010} (2010) 085,
  [\href{http://arxiv.org/abs/1008.4884}{{\tt arXiv:1008.4884}}].

\bibitem{Falkowski:2017pss}
A.~Falkowski, M.~Gonz\'{a}lez-Alonso, and K.~Mimouni, {\it {Compilation of
  low-energy constraints on 4-fermion operators in the SMEFT}},  {\em JHEP}
  {\bf 08} (2017) 123, [\href{http://arxiv.org/abs/1706.03783}{{\tt
  arXiv:1706.03783}}].

\bibitem{deBlas:2017xtg}
J.~de~Blas, J.~C. Criado, M.~Perez-Victoria, and J.~Santiago, {\it {Effective
  description of general extensions of the Standard Model: the complete
  tree-level dictionary}},  {\em JHEP} {\bf 03} (2018) 109,
  [\href{http://arxiv.org/abs/1711.10391}{{\tt arXiv:1711.10391}}].

\bibitem{Cirigliano:2009wk}
V.~Cirigliano, J.~Jenkins, and M.~Gonzalez-Alonso, {\it {Semileptonic decays of
  light quarks beyond the Standard Model}},  {\em Nucl. Phys.} {\bf B830}
  (2010) 95--115, [\href{http://arxiv.org/abs/0908.1754}{{\tt
  arXiv:0908.1754}}].

\bibitem{Cirigliano:2012ab}
V.~Cirigliano, M.~Gonzalez-Alonso, and M.~L. Graesser, {\it {Non-standard
  Charged Current Interactions: beta decays versus the LHC}},  {\em JHEP} {\bf
  02} (2013) 046, [\href{http://arxiv.org/abs/1210.4553}{{\tt
  arXiv:1210.4553}}].

\bibitem{Aebischer:2017gaw}
J.~Aebischer, M.~Fael, C.~Greub, and J.~Virto, {\it {B physics Beyond the
  Standard Model at One Loop: Complete Renormalization Group Evolution below
  the Electroweak Scale}},  \href{http://arxiv.org/abs/1704.06639}{{\tt
  arXiv:1704.06639}}.

\bibitem{Gonzalez-Alonso:2017iyc}
M.~Gonz\'{a}lez-Alonso, J.~Martin~Camalich, and K.~Mimouni, {\it
  {Renormalization-group evolution of new physics contributions to
  (semi)leptonic meson decays}},  {\em Phys. Lett.} {\bf B772} (2017) 777--785,
  [\href{http://arxiv.org/abs/1706.00410}{{\tt arXiv:1706.00410}}].

\bibitem{Lee:1956qn}
T.~D. Lee and C.-N. Yang, {\it {Question of Parity Conservation in Weak
  Interactions}},  {\em Phys. Rev.} {\bf 104} (1956) 254--258.

\bibitem{Ademollo:1964sr}
M.~Ademollo and R.~Gatto, {\it {Nonrenormalization Theorem for the Strangeness
  Violating Vector Currents}},  {\em Phys. Rev. Lett.} {\bf 13} (1964)
  264--265.

\bibitem{Gonzalez-Alonso:2018omy}
M.~Gonzalez-Alonso, O.~Naviliat-Cuncic, and N.~Severijns, {\it {New physics
  searches in nuclear and neutron $\beta$ decay}},  {\em Prog. Part. Nucl.
  Phys.} {\bf 104} (2019) 165--223,
  [\href{http://arxiv.org/abs/1803.08732}{{\tt arXiv:1803.08732}}].

\bibitem{Bhattacharya:2016zcn}
T.~Bhattacharya, V.~Cirigliano, S.~Cohen, R.~Gupta, H.-W. Lin, and B.~Yoon,
  {\it {Axial, Scalar and Tensor Charges of the Nucleon from 2+1+1-flavor
  Lattice QCD}},  {\em Phys. Rev.} {\bf D94} (2016), no.~5 054508,
  [\href{http://arxiv.org/abs/1606.07049}{{\tt arXiv:1606.07049}}].

\bibitem{Gonzalez-Alonso:2013ura}
M.~Gonz\'alez-Alonso and J.~Martin~Camalich, {\it {Isospin breaking in the
  nucleon mass and the sensitivity of $\beta$ decays to new physics}},  {\em
  Phys. Rev. Lett.} {\bf 112} (2014), no.~4 042501,
  [\href{http://arxiv.org/abs/1309.4434}{{\tt arXiv:1309.4434}}].

\bibitem{Giunti:2007ry}
C.~Giunti and C.~W. Kim, {\em {Fundamentals of Neutrino Physics and
  Astrophysics}}.
\newblock 2007.

\bibitem{Fierz1937}
M.~Fierz, {\it Zur fermischen theorie des $\beta$-zerfalls},  {\em Z. Phys.}
  {\bf 104} (1937), no.~7 553--565.

\bibitem{Mueller:2011nm}
T.~A. Mueller et~al., {\it {Improved Predictions of Reactor Antineutrino
  Spectra}},  {\em Phys. Rev.} {\bf C83} (2011) 054615,
  [\href{http://arxiv.org/abs/1101.2663}{{\tt arXiv:1101.2663}}].

\bibitem{Huber:2011wv}
P.~Huber, {\it {On the determination of anti-neutrino spectra from nuclear
  reactors}},  {\em Phys. Rev.} {\bf C84} (2011) 024617,
  [\href{http://arxiv.org/abs/1106.0687}{{\tt arXiv:1106.0687}}]. [Erratum:
  Phys. Rev.C85,029901(2012)].

\bibitem{King:1958zz}
R.~W. King and J.~F. Perkins, {\it {Inverse Beta Decay and the Two-Component
  Neutrino}},  {\em Phys. Rev.} {\bf 112} (1958) 963--966.

\bibitem{Davis:1979gg}
B.~R. Davis, P.~Vogel, F.~M. Mann, and R.~E. Schenter, {\it {Reactor
  anti-neutrino spectra and their application to anti-neutrino induced
  reactions}},  {\em Phys. Rev.} {\bf C19} (1979) 2259--2266.

\bibitem{Vogel:2007du}
P.~Vogel, {\it {Conversion of electron spectrum associated with fission into
  the antineutrino spectrum}},  {\em Phys. Rev.} {\bf C76} (2007) 025504,
  [\href{http://arxiv.org/abs/0708.0556}{{\tt arXiv:0708.0556}}].

\bibitem{Hayes:2016qnu}
A.~C. Hayes and P.~Vogel, {\it {Reactor Neutrino Spectra}},  {\em Ann. Rev.
  Nucl. Part. Sci.} {\bf 66} (2016) 219--244,
  [\href{http://arxiv.org/abs/1605.02047}{{\tt arXiv:1605.02047}}].

\bibitem{Glick-Magid:2016rsv}
A.~Glick-Magid, Y.~Mishnayot, I.~Mukul, M.~Hass, S.~Vaintraub, G.~Ron, and
  D.~Gazit, {\it {Beta spectrum of unique first-forbidden decays as a novel
  test for fundamental symmetries}},  {\em Phys. Lett.} {\bf B767} (2017)
  285--288, [\href{http://arxiv.org/abs/1609.03268}{{\tt arXiv:1609.03268}}].

\bibitem{Hayen:2018uyg}
L.~Hayen, J.~Kostensalo, N.~Severijns, and J.~Suhonen, {\it {First-forbidden
  transitions in reactor antineutrino spectra}},  {\em Phys. Rev.} {\bf C99}
  (2019), no.~3 031301, [\href{http://arxiv.org/abs/1805.12259}{{\tt
  arXiv:1805.12259}}].

\bibitem{Wells:2005vk}
J.~D. Wells, {\it {TASI lecture notes: Introduction to precision electroweak
  analysis}},  \href{http://arxiv.org/abs/hep-ph/0512342}{{\tt
  hep-ph/0512342}}.

\bibitem{Descotes-Genon:2018foz}
S.~Descotes-Genon, A.~Falkowski, M.~Fedele, M.~Gonz\'{a}lez-Alonso, and
  J.~Virto, {\it {The CKM parameters in the SMEFT}},  {\em JHEP} {\bf 05}
  (2019) 172, [\href{http://arxiv.org/abs/1812.08163}{{\tt arXiv:1812.08163}}].

\bibitem{Langacker:1988ur}
P.~Langacker and D.~London, {\it {Mixing Between Ordinary and Exotic
  Fermions}},  {\em Phys. Rev.} {\bf D38} (1988) 886.

\bibitem{Mention:2011rk}
G.~Mention, M.~Fechner, T.~Lasserre, T.~A. Mueller, D.~Lhuillier, M.~Cribier,
  and A.~Letourneau, {\it {The Reactor Antineutrino Anomaly}},  {\em Phys.
  Rev.} {\bf D83} (2011) 073006, [\href{http://arxiv.org/abs/1101.2755}{{\tt
  arXiv:1101.2755}}].

\bibitem{An:2016ses}
{\bf Daya Bay} Collaboration, F.~P. An et~al., {\it {Measurement of electron
  antineutrino oscillation based on 1230 days of operation of the Daya Bay
  experiment}},  {\em Phys. Rev.} {\bf D95} (2017), no.~7 072006,
  [\href{http://arxiv.org/abs/1610.04802}{{\tt arXiv:1610.04802}}].

\bibitem{Esmaili:2013yea}
A.~Esmaili, E.~Kemp, O.~L.~G. Peres, and Z.~Tabrizi, {\it {Probing light
  sterile neutrinos in medium baseline reactor experiments}},  {\em Phys. Rev.}
  {\bf D88} (2013) 073012, [\href{http://arxiv.org/abs/1308.6218}{{\tt
  arXiv:1308.6218}}].

\bibitem{Hardy:2014qxa}
J.~C. Hardy and I.~S. Towner, {\it {Superallowed $0^+\to 0^+$ nuclear $\beta$
  decays: 2014 critical survey, with precise results for $V_{ud}$ and CKM
  unitarity}},  {\em Phys. Rev.} {\bf C91} (2015), no.~2 025501,
  [\href{http://arxiv.org/abs/1411.5987}{{\tt arXiv:1411.5987}}].

\bibitem{Markisch:2018ndu}
B.~M{\"a}rkisch et~al., {\it {Measurement of the Weak Axial-Vector Coupling
  Constant in the Decay of Free Neutrons Using a Pulsed Cold Neutron Beam}},
  \href{http://arxiv.org/abs/1812.04666}{{\tt arXiv:1812.04666}}.

\bibitem{Tanabashi:2018oca}
{\bf Particle Data Group} Collaboration, M.~Tanabashi et~al., {\it {Review of
  Particle Physics}},  {\em Phys. Rev.} {\bf D98} (2018), no.~3 030001.

\bibitem{Marciano:2005ec}
W.~J. Marciano and A.~Sirlin, {\it {Improved calculation of electroweak
  radiative corrections and the value of V(ud)}},  {\em Phys. Rev. Lett.} {\bf
  96} (2006) 032002, [\href{http://arxiv.org/abs/hep-ph/0510099}{{\tt
  hep-ph/0510099}}].

\bibitem{Seng:2018yzq}
C.-Y. Seng, M.~Gorchtein, H.~H. Patel, and M.~J. Ramsey-Musolf, {\it {Reduced
  hadronic uncertainty in the determination of $V_{ud}$}},  {\em Phys. Rev.
  Lett.} {\bf 121} (2018), no.~24 241804,
  [\href{http://arxiv.org/abs/1807.10197}{{\tt arXiv:1807.10197}}].

\bibitem{Seng:2018qru}
C.~Y. Seng, M.~Gorchtein, and M.~J. Ramsey-Musolf, {\it {Dispersive Evaluation
  of the Inner Radiative Correction in Neutron and Nuclear $\beta$-decay}},
  \href{http://arxiv.org/abs/1812.03352}{{\tt arXiv:1812.03352}}.

\bibitem{Britton:1992pg}
D.~Britton, S.~Ahmad, D.~Bryman, R.~Burnbam, E.~Clifford, et~al., {\it
  {Measurement of the $\pi^+ \to e^+\nu$ branching ratio}},  {\em
  Phys.Rev.Lett.} {\bf 68} (1992) 3000--3003.

\bibitem{Czapek:1993kc}
G.~Czapek, A.~Federspiel, A.~Fluckiger, D.~Frei, B.~Hahn, et~al., {\it
  {Branching ratio for the rare pion decay into positron and neutrino}},  {\em
  Phys.Rev.Lett.} {\bf 70} (1993) 17--20.

\bibitem{Cirigliano:2007xi}
V.~Cirigliano and I.~Rosell, {\it {Two-loop effective theory analysis of $\pi
  (K) \to e \bar{\nu}_e (\gamma)$ branching ratios}},  {\em Phys.Rev.Lett.}
  {\bf 99} (2007) 231801, [\href{http://arxiv.org/abs/0707.3439}{{\tt
  arXiv:0707.3439}}].

\bibitem{Voloshin:1992sn}
M.~B. Voloshin, {\it {Upper bound on tensor interaction in the decay $\pi^- \to
  e^- \bar \nu \gamma$ }},  {\em Phys. Lett.} {\bf B283} (1992) 120--122.

\bibitem{Bhattacharya:2011qm}
T.~Bhattacharya, V.~Cirigliano, S.~D. Cohen, A.~Filipuzzi, M.~Gonzalez-Alonso,
  M.~L. Graesser, R.~Gupta, and H.-W. Lin, {\it {Probing Novel Scalar and
  Tensor Interactions from (Ultra)Cold Neutrons to the LHC}},  {\em Phys. Rev.}
  {\bf D85} (2012) 054512, [\href{http://arxiv.org/abs/1110.6448}{{\tt
  arXiv:1110.6448}}].

\bibitem{Gupta:2018qil}
R.~Gupta, Y.-C. Jang, B.~Yoon, H.-W. Lin, V.~Cirigliano, and T.~Bhattacharya,
  {\it {Isovector Charges of the Nucleon from 2+1+1-flavor Lattice QCD}},  {\em
  Phys. Rev.} {\bf D98} (2018) 034503,
  [\href{http://arxiv.org/abs/1806.09006}{{\tt arXiv:1806.09006}}].

\bibitem{Aaboud:2017efa}
{\bf ATLAS} Collaboration, M.~Aaboud et~al., {\it {Search for a new heavy gauge
  boson resonance decaying into a lepton and missing transverse momentum in 36
  fb$^{-1}$ of $pp$ collisions at $\sqrt{s} =$ 13 TeV with the ATLAS
  experiment}},  {\em Eur. Phys. J.} {\bf C78} (2018), no.~5 401,
  [\href{http://arxiv.org/abs/1706.04786}{{\tt arXiv:1706.04786}}].

\bibitem{Pruna:2015jhf}
G.~M. Pruna and A.~Signer, {\it {Lepton-flavour violating decays in theories
  with dimension 6 operators}},  {\em EPJ Web Conf.} {\bf 118} (2016) 01031,
  [\href{http://arxiv.org/abs/1511.04421}{{\tt arXiv:1511.04421}}].

\bibitem{Crivellin:2017rmk}
A.~Crivellin, S.~Davidson, G.~M. Pruna, and A.~Signer, {\it
  {Renormalisation-group improved analysis of $\mu\to e$ processes in a
  systematic effective-field-theory approach}},  {\em JHEP} {\bf 05} (2017)
  117, [\href{http://arxiv.org/abs/1702.03020}{{\tt arXiv:1702.03020}}].

\bibitem{Davidson:2018rqt}
S.~Davidson and A.~Saporta, {\it {Constraints on $2\ell 2q$ operators from
  $\mu$-$e$ flavour-changing meson decays}},  {\em Phys. Rev.} {\bf D99}
  (2019), no.~1 015032, [\href{http://arxiv.org/abs/1807.10288}{{\tt
  arXiv:1807.10288}}].

\bibitem{Coy:2018bxr}
R.~Coy and M.~Frigerio, {\it {Effective approach to lepton observables: the
  seesaw case}},  \href{http://arxiv.org/abs/1812.03165}{{\tt
  arXiv:1812.03165}}.

\bibitem{Cirigliano:2009bz}
V.~Cirigliano, R.~Kitano, Y.~Okada, and P.~Tuzon, {\it {On the model
  discriminating power of $mu \to e$ conversion in nuclei}},  {\em Phys. Rev.}
  {\bf D80} (2009) 013002, [\href{http://arxiv.org/abs/0904.0957}{{\tt
  arXiv:0904.0957}}].

\bibitem{Bertl:2006up}
{\bf SINDRUM II} Collaboration, W.~H. Bertl et~al., {\it {A Search for muon to
  electron conversion in muonic gold}},  {\em Eur. Phys. J.} {\bf C47} (2006)
  337--346.

\bibitem{Celis:2014asa}
A.~Celis, V.~Cirigliano, and E.~Passemar, {\it {Model-discriminating power of
  lepton flavor violating $\tau$ decays}},  {\em Phys. Rev.} {\bf D89} (2014),
  no.~9 095014, [\href{http://arxiv.org/abs/1403.5781}{{\tt arXiv:1403.5781}}].

\end{thebibliography}\endgroup

\end{document}